\documentclass[%
 reprint,
 amsmath,amssymb,
 aps,
pra,
floatfix,
]{revtex4-2}

\usepackage{graphicx}
\usepackage{dcolumn}
\usepackage{bm}
\usepackage{amsmath,amssymb}
\usepackage{mathtools}

\usepackage{tikz}

\usepackage{multirow}
\usepackage{array}

\usepackage[colorlinks=true,bookmarks=false,citecolor=blue,urlcolor=blue]{hyperref}

\begin{document}

\title{Stochastic Dynamics of Domain Walls on a Racetrack: Impact of Inhomogeneity of Magnetic Anisotropy}

\author{Anton~V.~Hlushchenko$^1$}
\author{Mykhailo~I.~Bratchenko$^1$}
\author{Oksana~L.~Andrieieva$^1$}
\author{Kostyantyn~I.~Polozhiy$^{2}$}
\author{Andriy~M.~Styervoyedov$^{2}$}
\author{Igor M. Sokolov$^3$}
\author{Samudrajit~Thapa$^{4,5}$}
\author{Stuart~S.~P.~Parkin$^2$}
\author{Aleksei~V.~Chechkin$^{1,2,5,6}$}
 \email{achechkin@mpi-halle.mpg.de}
\affiliation{$^1$National Science Center "Kharkiv Institute of Physics and Technology", 61108 Kharkiv, Ukraine}
\affiliation{$^2$Max Planck Institute of Microstructure Physics, 06120 Halle, Germany}
\affiliation{$^3$Institut für Physik, Humboldt-Universität zu Berlin, Newtonstraße 15, D-12489 Berlin, Germany}
\affiliation{$^4$Department of Physics, Indian Institute of Technology Guwahati, Guwahati 781039, Assam, India}
\affiliation{$^5$Asia Pacific Centre for Theoretical Physics, Pohang 37673, Republic of Korea}
\affiliation{$^6$Faculty of Pure and Applied Mathematics, Wrocław University of Science and Technology, 50-370 Wrocław, Poland}

\author
\author

\date{\today}

\begin{abstract}

Racetrack memory is a spintronic technology that stores information in current-driven magnetic domain walls (DW), offering a promising route toward non-volatile, high-density, and high-speed memory devices. In ultrathin magnetic films, atomic-scale interface disorder can produce spatially inhomogeneous magnetic anisotropy, creating a complex spatially varying energy landscape that strongly influences DW motion.
In this work, we model anisotropy inhomogeneities using a two-dimensional fractal Weierstrass--Mandelbrot function, which provides independent control over the disorder amplitude and characteristic spatial scale. Micromagnetic simulations reveal pronounced stochasticity of current-driven domain-wall transport even in the absence of thermal fluctuations. Transport is quantified by the transmission fraction, defined as the fraction of DWs that successfully reach a detector after propagating through a disordered racetrack. We investigate its dependence on current density, disorder amplitude, and the spatial scale of the anisotropy inhomogeneity. The resulting transmission statistics are described by Gaussian error-function profiles and exhibit a universal scaling behavior upon normalization by the corresponding median values.
Statistical analysis of DW trajectories further characterizes the disorder-induced fluctuations of the propagation process. The variance of the random DW positions shows a pronounced dependence on disorder amplitude and is well described by the proposed phenomenological model.
These results provide a quantitative framework for describing stochastic DW transport in structurally disordered racetracks and establish disorder-induced reliability limits in racetrack-memory devices.


\end{abstract}

\maketitle

\section{Introduction}

Racetrack memory is a promising spintronic technology in which information is stored in the form of magnetic textures, such as domain walls (DWs) and skyrmions, that are driven along nanoscale magnetic tracks by electric currents \cite{Parkin_2008,Parkin2015,Fert2013}. Owing to its non-volatility, high storage density, fast operation, and potential for three-dimensional integration, racetrack memory is considered a strong candidate for next-generation memory and logic architectures \cite{Vedmedenko2020,Parkin_2020,Gu2022,Farinha2025}. Continued advances in materials engineering and magnetic-texture manipulation are essential for realizing reliable and scalable racetrack-memory devices \cite{JaeChun_Science_2024}.

One of the major challenges in racetrack memory is the stochastic nature of magnetic-texture motion \cite{Nam_APL_2018,Lee2023,Ishibashi_SciAdv_2024}. In practical devices, densely packed spin textures must be displaced and detected with high precision \cite{JaeChun_ACSNANO_2024}. Thermal fluctuations constitute an unavoidable source of stochasticity in magnetic systems, while nonequilibrium spin-current noise associated with spin-transfer and spin-orbit torque mechanisms can further influence the dynamics of magnetic textures \cite{Tserkovnyak_shot_2005,Chudnovskiy_shot_2014,Thiaville_2018,Hlushchenko_2026}. 

In realistic magnetic materials, structural and magnetic inhomogeneities create a spatially varying energy landscape that can strongly modify the transport of magnetic textures, in addition to thermal and current-induced fluctuations \cite{Jiang2010,Lee_2011}. Geometrical modulations of magnetic nanowires and spatial variations of magnetic anisotropy have also been shown to affect domain-wall pinning and magnetic-texture dynamics \cite{ChubykaloFesenko2018,ChubykaloFesenko2023,Mertig2023}. Such quenched disorder gives rise to pinning, depinning transitions, threshold currents, and velocity fluctuations, resulting in stochastic domain-wall propagation \cite{Paul_1982_JApplPhys,Huse_1985_PhysRevLett,Rosso_2007_PhysRevB,Xi2015,Glatz_2018_PhysRevLett,Koyama2011,Hayward2015}.

A common approach to modeling disorder-induced pinning is to represent structural inhomogeneities as discrete pinning centers. Within this framework, the dynamics of domain walls has been studied in terms of the density and spatial distribution of pinning sites, with the characteristic distance between pinning centers relative to the domain-wall width playing an important role in determining the transport properties \cite{Gorchon_PRL_2014,Jeudy_PRL_2016,Jeudy_prb_2018,Thiaville_2020_PhysRevResearch}. Other approaches account for more extended or granular forms of disorder, providing descriptions of inhomogeneities specific to particular magnetic material  \cite{Castell_2023,Domenichini_2024_PhysRevB}. For skyrmions, however, the discrete-pinning-center picture may not adequately describe the experimentally observed transport behavior. It was demonstrated in Ref.~\cite{woo2016observation} that short-scale spatial variations of the energy landscape can strongly pin skyrmions, resulting in finite critical currents, reduced velocities, and current-induced annihilation of pinned textures. This work highlighted the importance of disorder length scales comparable to the characteristic size of a magnetic texture, motivating a multiscale description of the disorder landscape.

These results motivate describing disorder as a spatially distributed energy landscape rather than solely as a collection of isolated pinning centers. Here, we model the spatial variation of magnetic anisotropy using a random Weierstrass--Mandelbrot (W-M) function \cite{Berry1980}. The multivariate generalization proposed in Ref.~\cite{Ausloos_W_M_1985} enables the construction of multiscale random fields and has found applications in modeling self-affine rough surfaces, heterogeneous media, and spatially varying material properties \cite{Yan1998,Komvopoulos2000,PATRIKAR_2004,FENG2018489,Yu_Fractal_2018,Papanikolaou2019,Hlushchenko_inhomogeneity_24,Hlushchenko_inhomogeneity_25}. In our model, the resulting anisotropy variations generate an effective disordered energy landscape governing domain-wall dynamics. The W-M formalism allows the disorder amplitude and characteristic spatial scale to be varied independently, enabling a systematic investigation of their roles in current-driven domain-wall transport.

In this work, we investigate stochastic current-driven domain-wall transport in racetracks with spatially inhomogeneous magnetic anisotropy using large-scale micromagnetic simulations. We quantify transport by the transmission fraction, i.e., the fraction of domain walls that successfully reach a detector after propagating through a disordered racetrack, and determine how this quantity depends on the amplitude and characteristic scale of the anisotropy disorder. We demonstrate that the transmission curves are accurately described by a Gaussian error-function profile and exhibit universal scaling after normalization by the median current density. Furthermore, by analyzing the statistical properties of domain-wall trajectories, we reveal a pronounced non-monotonic dependence of trajectory fluctuations on the disorder amplitude. These results establish a quantitative framework for understanding stochastic domain-wall transport in structurally disordered magnetic racetracks and provide practical guidelines for evaluating disorder-induced reliability limits in racetrack-based spintronic devices.

\begin{figure*}
\centering
\includegraphics[width=\linewidth]{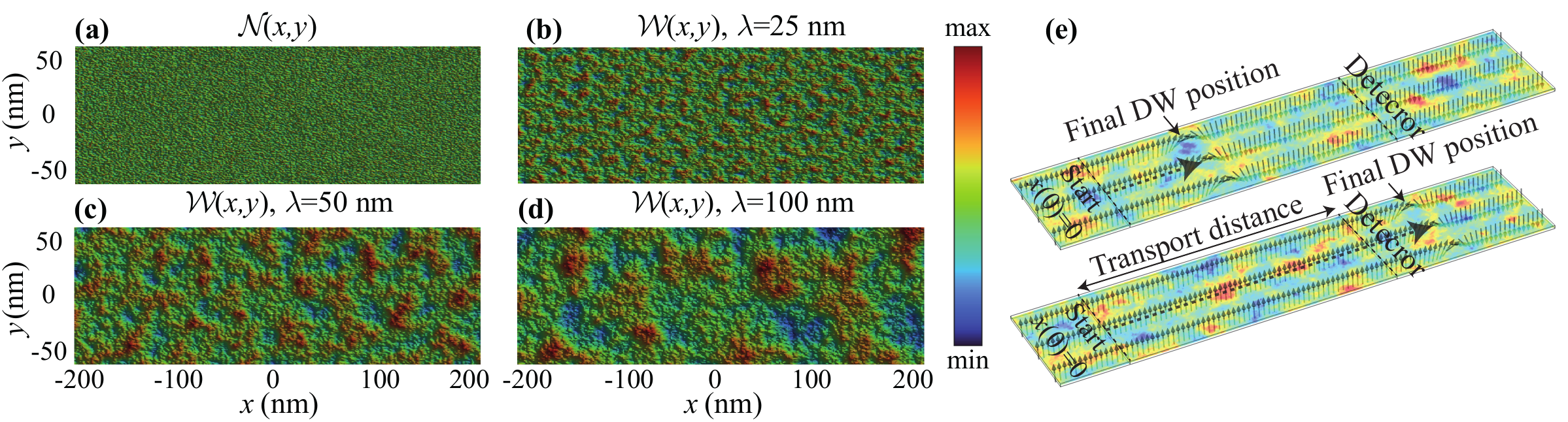}
\includegraphics[width=5cm]{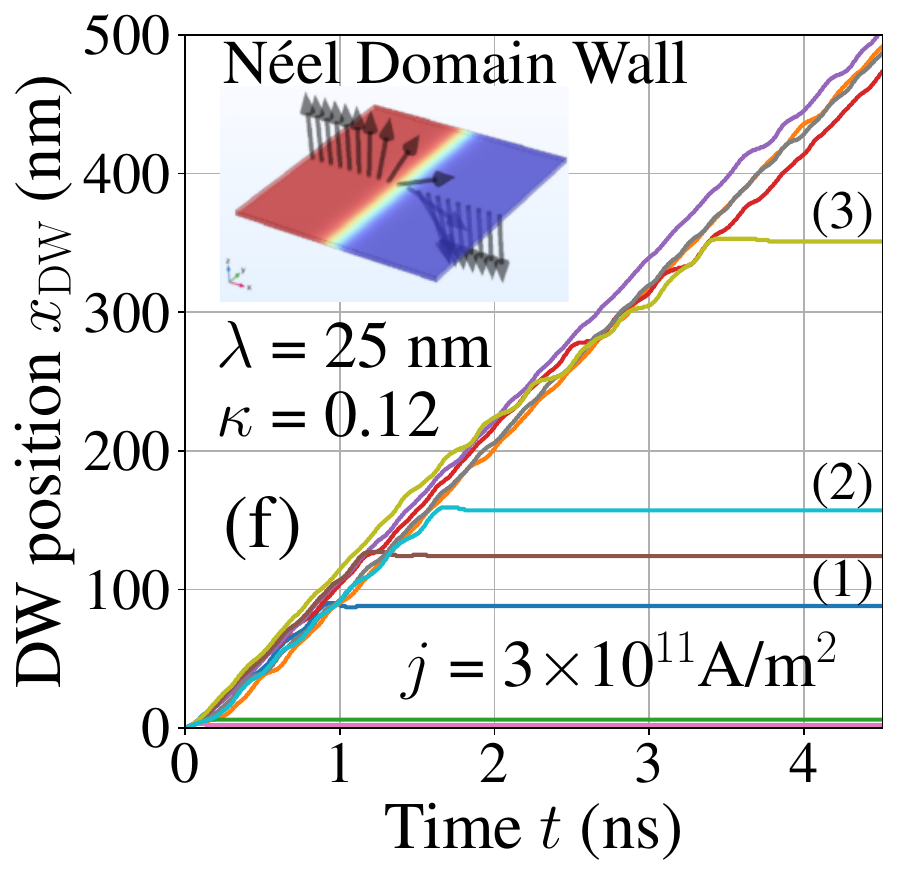}
\includegraphics[width=12cm]{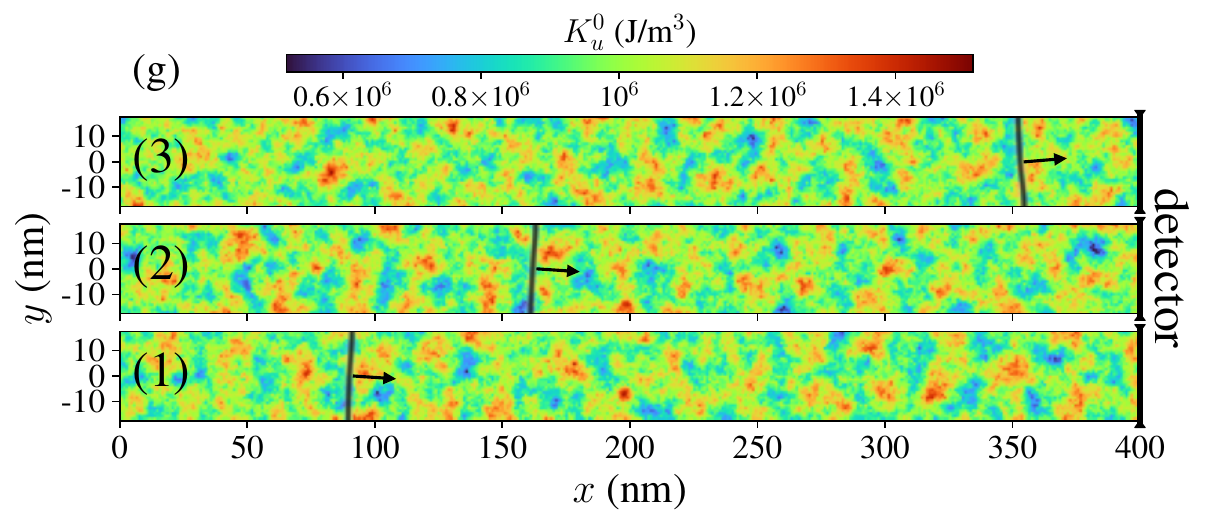}
\caption{\label{Displacements}
(a) White-noise field $\mathcal{N}(x,y)$ and (b)–(d) representative realizations of the Weierstrass--Mandelbrot random function $\mathcal{W}(x,y)$ for different values of the scale factor $\lambda$. (e) Schematic illustration of two representative domain-wall transport outcomes in racetracks with inhomogeneous magnetic anisotropy: the upper realization shows a domain wall pinned before reaching the detector, whereas the lower realization corresponds to successful transmission through the detector. (f) Representative domain-wall trajectories for several disorder realizations at $j=3\times10^{11}$ A/m$^2$, $\kappa=0.12$, and $\lambda=25$ nm. (g) Magnetic-anisotropy landscapes corresponding to the trajectories labeled (1)–(3) in (f), with arrows indicating the final domain-wall positions after $t=4.5$ ns.
}
\end{figure*}

\section{Inhomogeneous Anisotropy in Micromagnetic Simulations}
\begin{figure*}
\centering
\includegraphics[width=5.9cm]{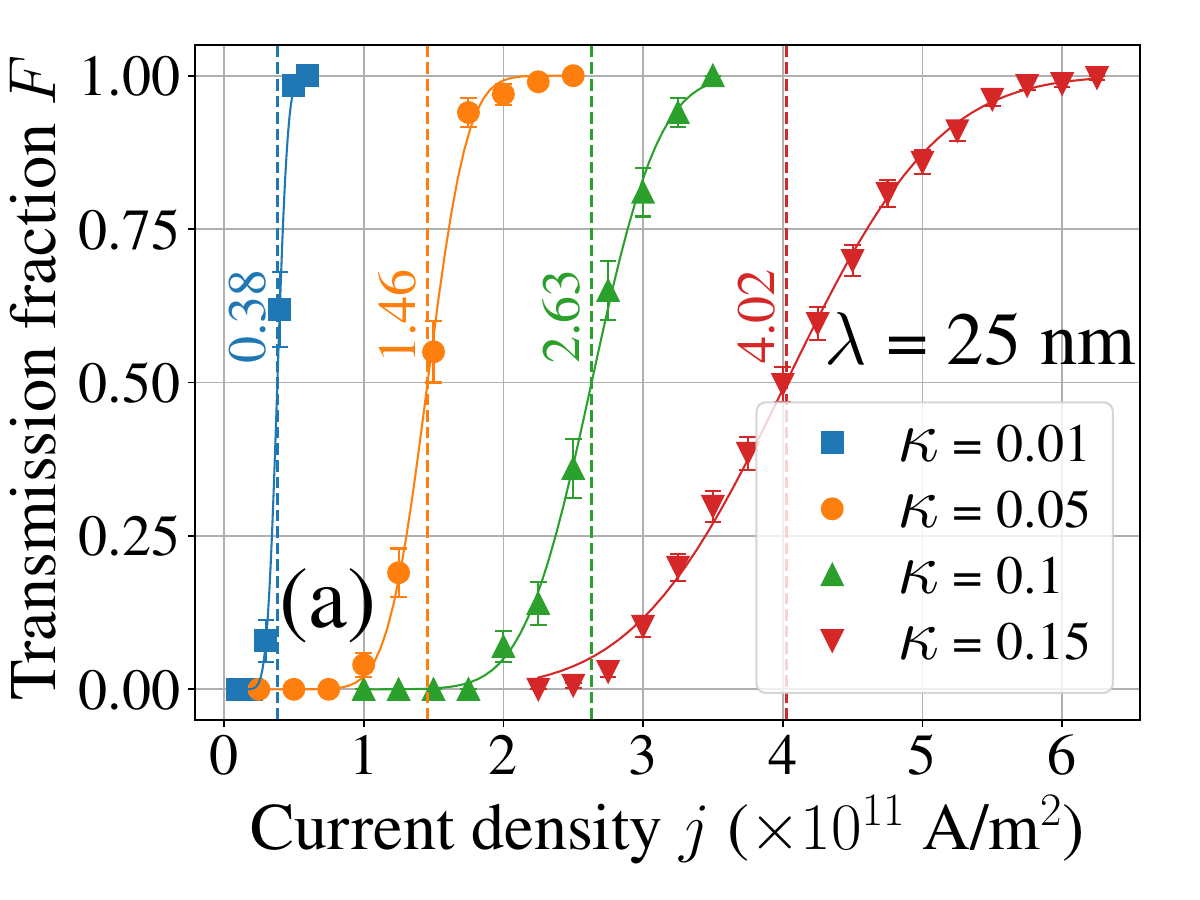}
\includegraphics[width=5.9cm]{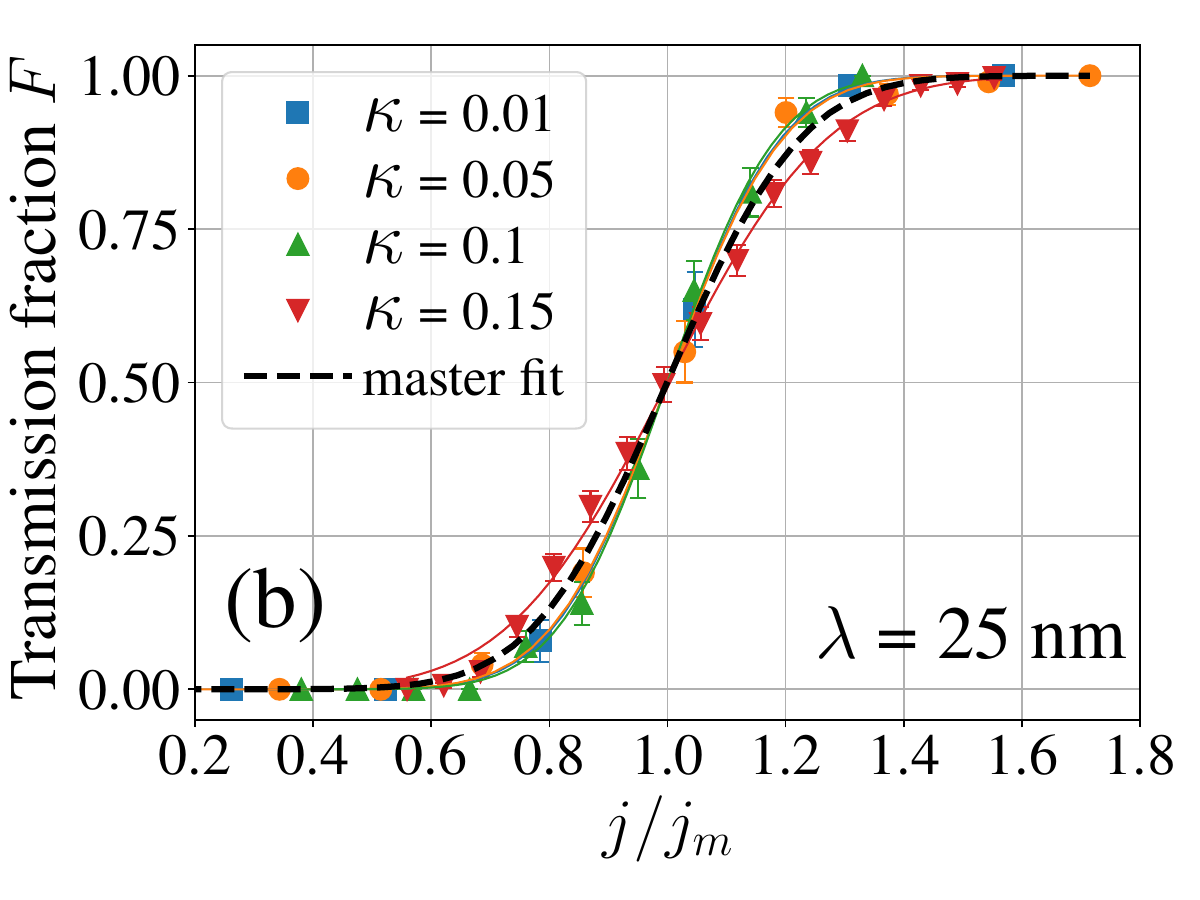}
\includegraphics[width=5.9cm]{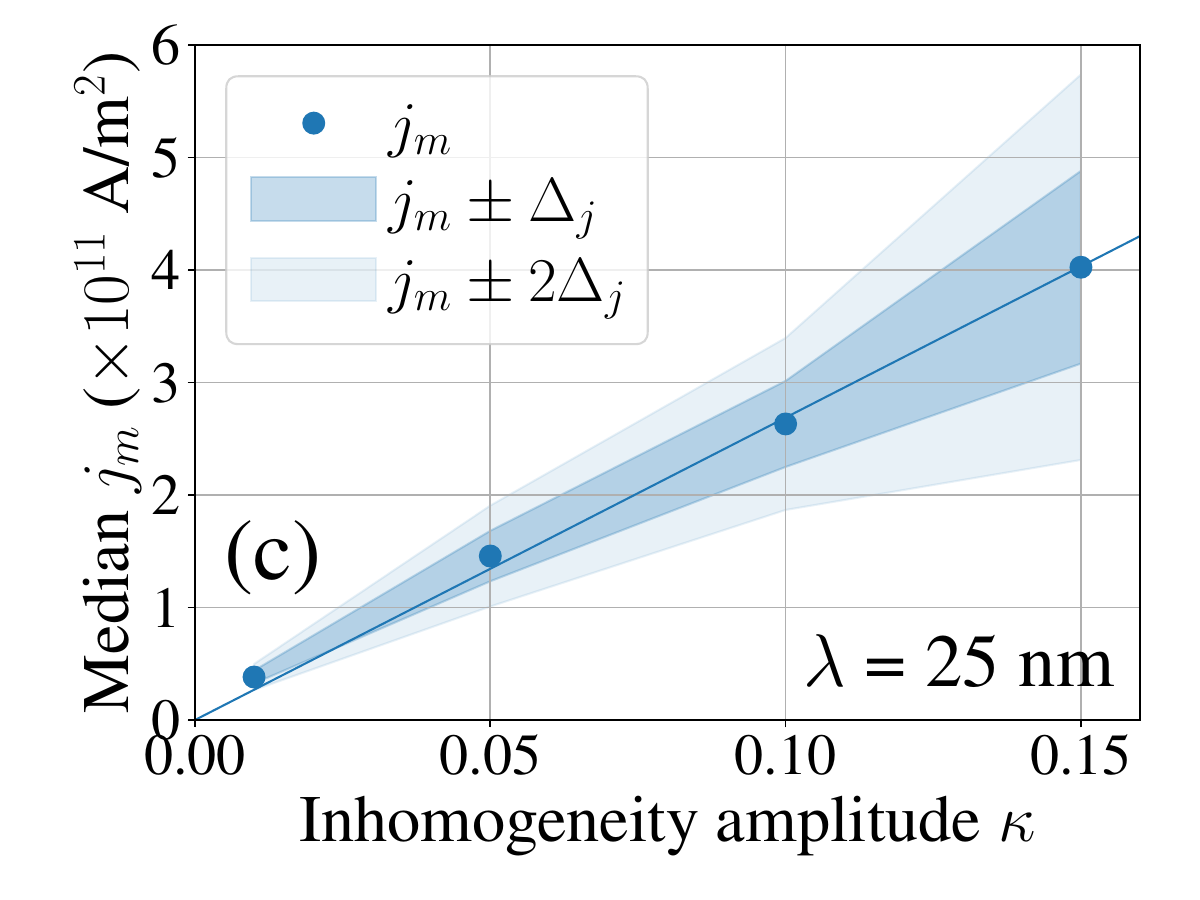}
\includegraphics[width=5.9cm]{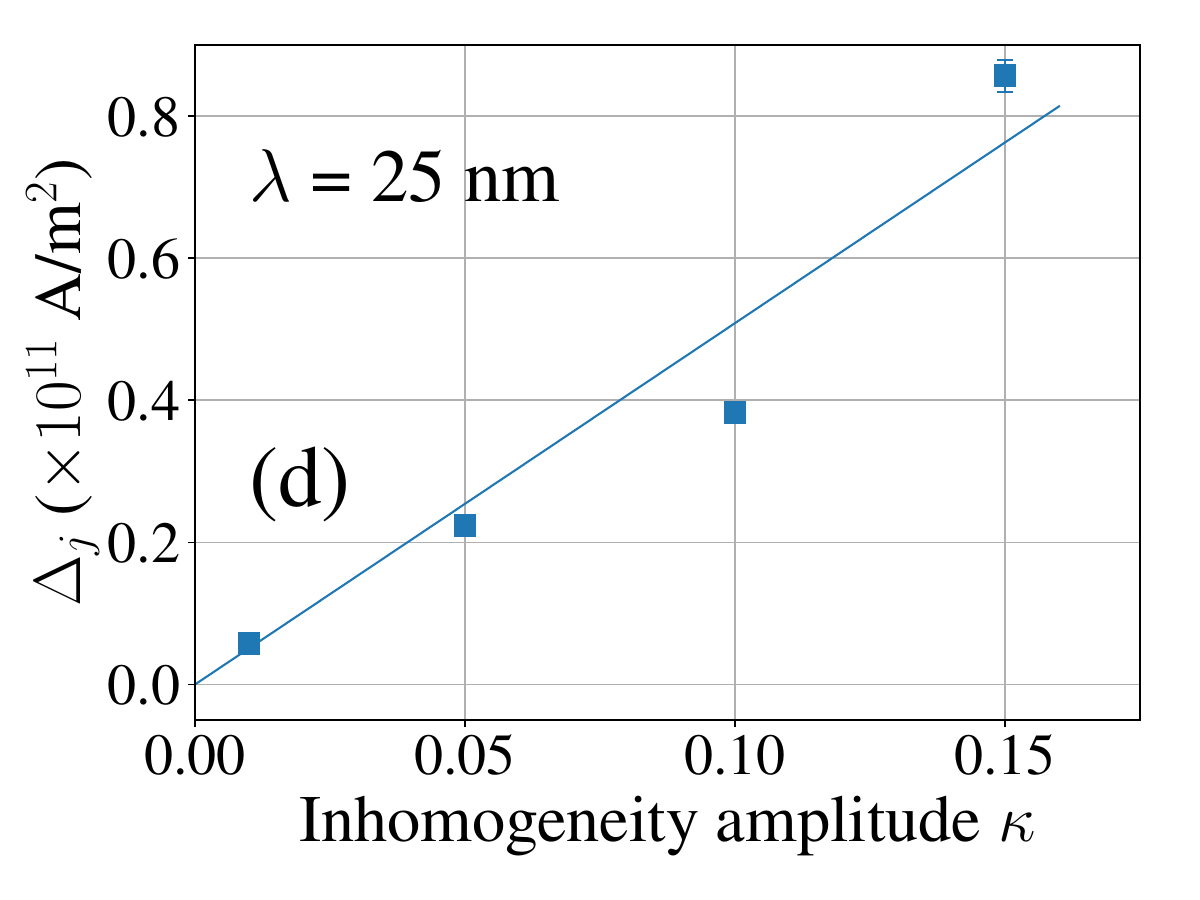}
\includegraphics[width=5.9cm]{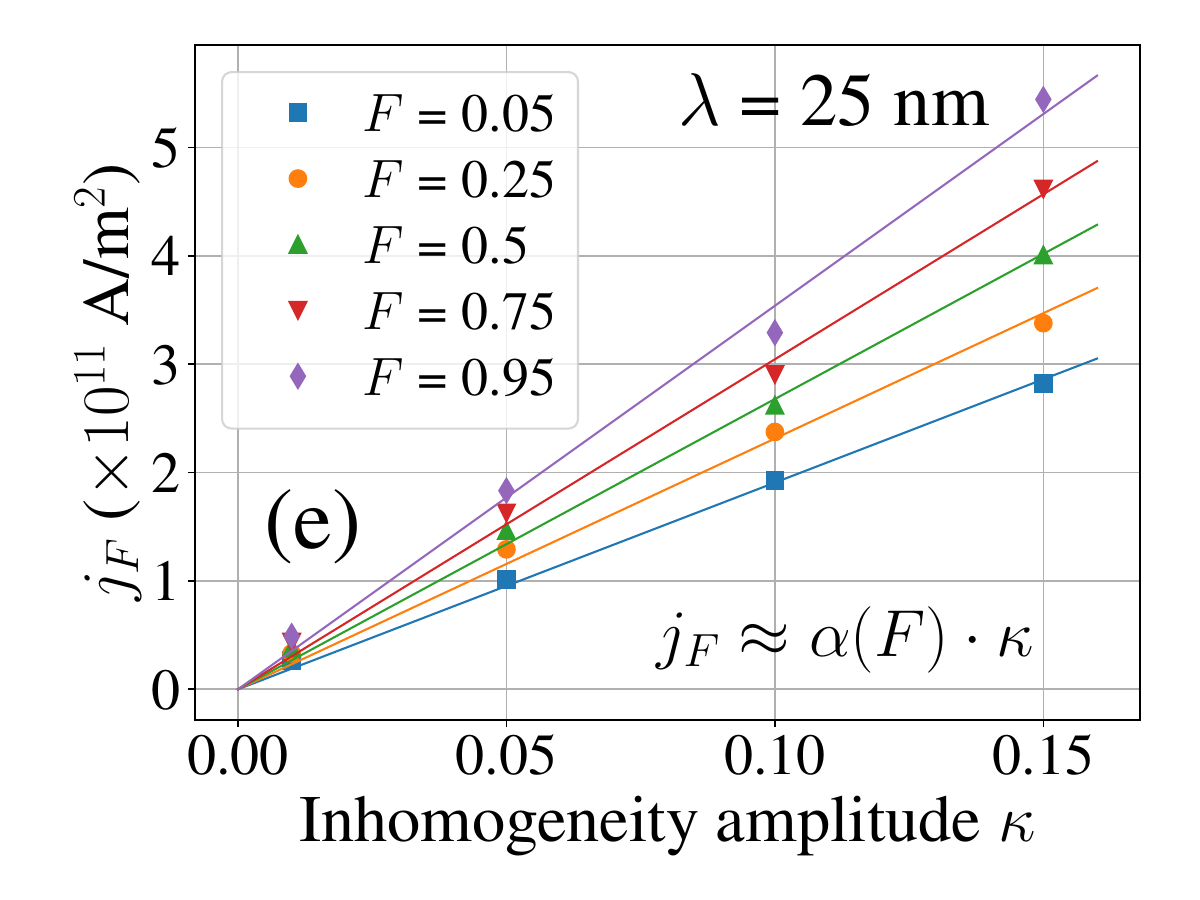}
\includegraphics[width=5.9cm]{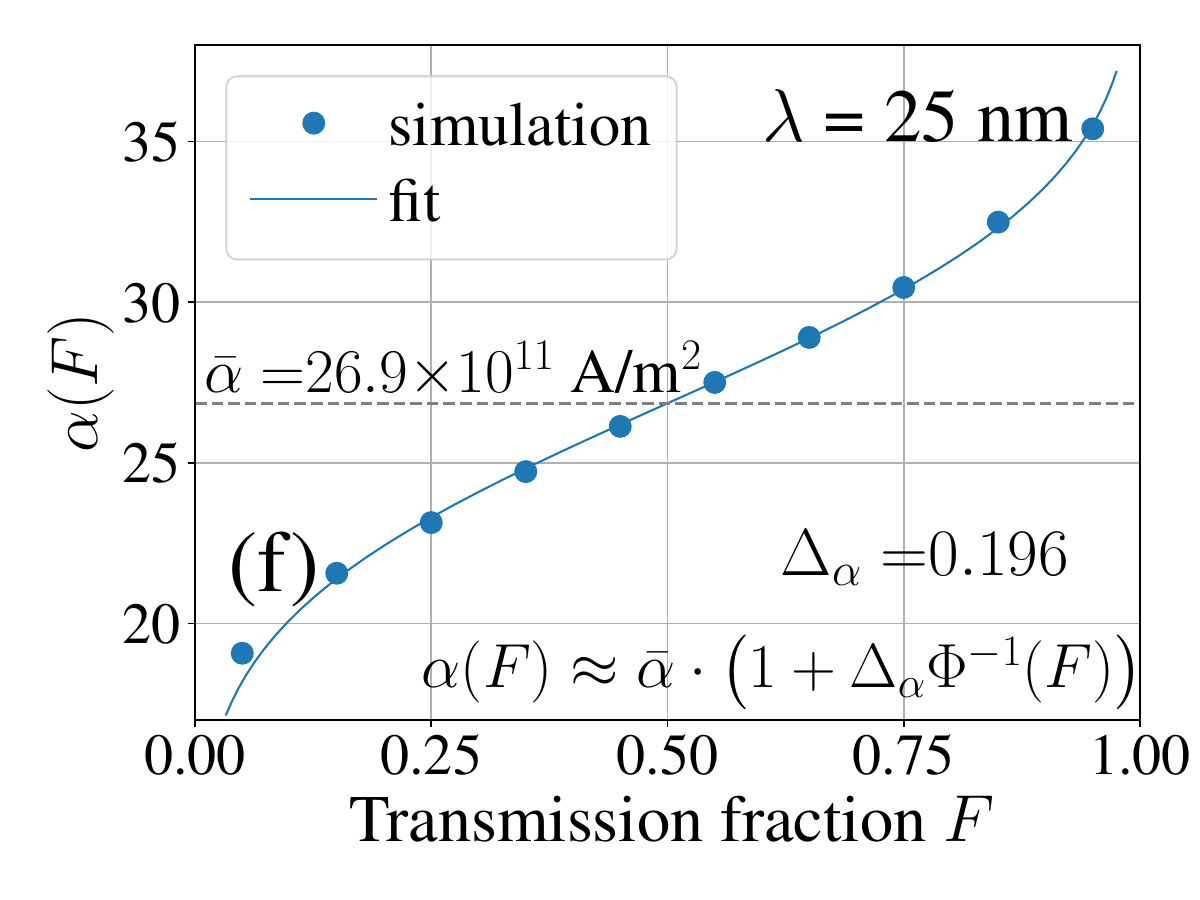}
\caption{\label{depin_prob}
(a) Transmission fraction $F$ as a function of the current density $j$ for different anisotropy inhomogeneity amplitudes $\kappa$ at the characteristic scale $\lambda=25$~nm. The numbers shown next to each curve indicate the corresponding median current density $j_m$. Symbols represent micromagnetic simulation results, while solid lines are fits by a Gaussian error function, $\Phi\left(j\right)$. (b) The same data plotted as a function of the reduced current density $j/j_m$. The dashed black line shows the master curve obtained from the Gaussian error-function fit. (c) Median current density $j_m$ as a function of the anisotropy inhomogeneity amplitude $\kappa$. The shaded regions indicate the intervals $j_m\pm\Delta_j$ and $j_m\pm2\Delta_j$, where transmission fraction changes the most. (d) Transition width $\Delta_j$ as a function of the anisotropy inhomogeneity amplitude $\kappa$. (e) Current density $j_F$ required to achieve prescribed transmission fractions $F$ as a function of the anisotropy inhomogeneity amplitude $\kappa$. Solid lines represent linear fits $j_F\approx \alpha(F)\kappa$. (f) Fitted slopes $\alpha(F)$ extracted from the linear dependencies shown in panel (e) as a function of the transmission fraction $F$. The solid line represents a fit by the inverse Gaussian error function, $\Phi^{-1}(F)$.}
\end{figure*}

The magnetization dynamics is described by the Landau-Lifshitz-Gilbert (LLG) equation \cite{Landau_PhysSowjet_1935,Gilbert_IEEE_2004} supplemented by the spin-orbit torque (SOT) term \cite{Slonczewski_1996},
\begin{equation}
\begin{split}\label{LLG_torque}
\partial_t\textbf{m}=
-\gamma\textbf{m}\times\textbf{H}_\text{eff}+\alpha\textbf{m}\times\partial_t\textbf{m}+\textbf{T}_\text{SOT}.
\end{split}
\end{equation}
where $\mathbf{m}(\mathbf{r},t)=\mathbf{M}(\mathbf{r},t)/M_s$ is the magnetization normalized by the saturation magnetization $M_s$, $\mathbf{M}$ is the local magnetization vector, $\mathbf{H}_{\rm eff}$ is the effective magnetic field, $\gamma$ is the gyromagnetic ratio, and $\alpha$ is the Gilbert damping constant.
The spin-orbit torque $\textbf{T}_\text{SOT}$ is written as
\begin{equation}
\begin{split}\label{Slonczewski}
\textbf{T}_\text{SOT}=
-\frac{\gamma \hbar P j}{2 e M_s \Delta_f} \, \mathbf{m} \times \left( \mathbf{m} \times \mathbf{p} \right)
-\frac{\gamma \beta \hbar P j}{2 e M_s \Delta_f} \, \mathbf{m} \times \mathbf{p},
\end{split}
\end{equation}
which depends on spin polarization $P$, current density $j$, reduced Planck constant $\hbar$, elementary charge $e$, and ferromagnetic layer thickness $\Delta_f$. The parameter $\beta$ controls the strength of the damping-like term, while
$\mathbf{p}=(\cos\psi,\sin\psi,0)$
defines the spin-polarization direction. No thermal fluctuations are included in the simulations, allowing us to isolate the influence of structural disorder on domain-wall dynamics. 

The effective field entering Eq.~(\ref{LLG_torque}) contains exchange, perpendicular anisotropy, and interfacial Dzyaloshinskii-Moriya interaction (DMI) contributions,
\begin{equation}
\begin{split}\label{effective_field_mumax}
\textbf{H}_\text{eff}=\frac{2A_\text{ex}}{M_s}\nabla^2\textbf{m}+\frac{2K_u(x,y)}{M_s}m_z\textbf{e}_z+ \\ \frac{2D_\text{DMI}}{M_s}(\nabla{m}_z-(\nabla\cdot\textbf{m})\textbf{e}_z).
\end{split}
\end{equation}
Here $A_{\rm ex}$ is the exchange stiffness, $K_u(x,y)$ is the spatially varying perpendicular magnetic anisotropy, and $D_{\rm DMI}$ is the interfacial DMI constant. The dipole–dipole interactions can be neglected because the extreme aspect ratio of the racetrack reduces in-plane demagnetizing factors to near-zero, while the out-of-plane component is already integrated into the effective anisotropy field \cite{Mulkers_PhysRevB_2017,Gonzalez_PhysRevApplied_2022,Coey_2010}.

To model structural disorder, we assume spatial fluctuations of the anisotropy energy,
\begin{equation}
\begin{split}\label{anisotropy}
&K_{u}(x,y)=K_u^0+\kappa K_u^0\left(\mathcal{W}(x,y)-\overline{ \mathcal{W}(x,y)}\right),  \\
&\overline{ \mathcal{W}(x,y)}=\frac{1}{S}\int_S{\mathcal{W}(x,y)dxdy},
\end{split}
\end{equation}
where $K_u^0$ is the mean anisotropy constant, the bar on $\overline{ \mathcal{W}(x,y)}$ denotes average over the racetrack area  $S$ in the $(x,y)$ plane, and $\kappa$ controls the strength of the disorder.
The anisotropy landscape is generated by the random Weierstrass-Mandelbrot (W-M) function \cite{Ausloos_W_M_1985,Saupe_1988_fractal},
\begin{equation}
\begin{aligned}
&\mathcal{W}(x,y)=\frac{1}{\sigma_\mathcal{W}}\sum_{n=0}^{N-1} 
\sum_{m=1}^{M}\gamma_\mathcal{W}^{-n(3-D)}\times \\
&\cos{\left( k_n\left(x\cos{\theta_m} + y\sin{\theta_m}\right)-\phi(m,n)\right)}, 
\label{W_M}
\end{aligned}
\end{equation}
where
\begin{equation}
\begin{aligned}
&k_n=\frac{2\pi\gamma_\mathcal{W}^n}{\lambda}, \theta_m=\frac{2\pi m}{M}, \\ &\sigma_\mathcal{W} = \sqrt{\frac{M}{2}\frac{1-\gamma_\mathcal{W}^{-2(3-D)N}}{1-\gamma_\mathcal{W}^{-2(3-D)}}}. 
\label{W_M2}
\end{aligned}
\end{equation}
Here $\phi(m,n)$ are independent random phases uniformly distributed in the interval $[0,2\pi)$, $D$ $(2<D<3)$ is the fractal dimension, $\gamma_\mathcal{W}>1$ is the geometric progression factor, $\lambda$ is the characteristic scale of the anisotropy landscape, and $M$ and $N$ denote the numbers of terms in the angular and scale expansions, respectively. Unless stated otherwise, the simulations are performed with $\gamma_\mathcal{W}=1.5$, $M=N=25$, and $D=2.5$, as a conventional practical choice \cite{Yan1998,Komvopoulos2000,PATRIKAR_2004,FENG2018489,Yu_Fractal_2018,Papanikolaou2019,Hlushchenko_inhomogeneity_24,Hlushchenko_inhomogeneity_25}. The first two statistical moments of the W-M function satisfy $\langle \mathcal{W}(x,y)\rangle =0$ and $\langle \mathcal{W}^2(x,y)\rangle =1$, where the angular brackets denote ensemble-average over the disordered landscape, 
\begin{equation}
    \langle ...\rangle = \frac{1}{2\pi}\int_{0}^{2\pi}... \; d\phi. \label{ensemble}
\end{equation}
Our choice of the normalization factor $\sigma_\mathcal{W}$ ensures unit
variance of the W--M function (see Appendix~\ref{App_RMS} for its
analytical derivation). The parameter $\lambda$ defines the characteristic
scale factor of the disorder and is linearly proportional to the correlation
length $\mathcal{L}$, making $\lambda$ a convenient measure of the spatial
scale of anisotropy inhomogeneities (see Appendix~\ref{App_Corr_length1} for the
derivation of the correlation length).

To illustrate the disorder model introduced above, Fig.~\ref{Displacements}(a)--(d) shows representative realizations of the spatially varying anisotropy fields. Panel (a) presents the underlying white-noise field $\mathcal{N}(x,y)$, while panels (b)--(d) show the corresponding Weierstrass--Mandelbrot functions $\mathcal{W}(x,y)$ for different values of the scale factor $\lambda$. Increasing $\lambda$ produces progressively larger correlated regions in the disorder landscape.

To investigate the influence of this disorder on domain-wall transport, we performed micromagnetic simulations using \texttt{mumax}$+$ \cite{Moreels2026} of a Néel domain wall driven by spin-orbit torque in a perpendicularly magnetized racetrack. The racetrack dimensions are $500\times35\times1$~nm$^3$. The thickness of racetrack is $\Delta_f=1$~nm.  The material parameters correspond to a Pt/Co-based ultrathin film \cite{JaeChun_ACSNANO_2024,Hlushchenko_2026}, with exchange stiffness $A_{\rm ex}=10^{-11}$~J/m, average anisotropy constant $K_u^0=10^6$~J/m$^3$, saturation magnetization $M_s=0.6\times10^6$~A/m, interfacial DMI constant $D_{\rm DMI}=0.5\times10^{-3}$~J/m$^2$, and damping parameters $\alpha=\beta=0.3$. The spin-polarization direction is chosen as $\mathbf{p}=(\cos\psi,\sin\psi,0)$ with $\psi=\pi/2$ and $P=0.35$. 

Figure~\ref{Displacements}(e) schematically illustrates the transport geometry, showing a domain wall driven along the racetrack toward a detector. Depending on the realization of the anisotropy landscape, the domain wall may either become pinned before reaching the detector or propagate through it. Representative domain-wall trajectories for several disorder realizations are shown in Fig.~\ref{Displacements}(f) at a fixed current density $j=3\times10^{11}$~A/m$^2$, disorder amplitude $\kappa=0.12$, and scale factor $\lambda=25$~nm.

The corresponding anisotropy landscapes are shown in Fig.~\ref{Displacements}(g), where the arrows indicate the local magnetization direction at the final simulation time $t=4.5$~ns. Some realizations allow the domain wall to propagate over the full racetrack and reach the detector, whereas others strongly hinder its motion and lead to pinning at different positions. Despite the identical driving conditions, the domain walls exhibit substantially different propagation histories, reflecting the stochasticity induced by the quenched disorder. In several realizations, pronounced transient slowdowns are observed before the walls resume their motion and eventually reach the detector. Since thermal fluctuations are absent in the present simulations, this delayed propagation originates solely from the intrinsic dynamics of the extended domain walls interacting with the disorder landscape. The observed transient slowdowns are consistent with the established picture of domain walls as extended elastic objects interacting with quenched disorder, where the competition between domain-wall elasticity and disorder-induced pinning governs the depinning dynamics \cite{Gorchon_PRL_2014,Thiaville_2020_PhysRevResearch}. 

\section{Transmission Statistics in Inhomogeneous Racetracks}

In this section we study statistical characteristics of the current-driven domain wall transport. For each set of disorder parameters, statistical averages are evaluated over an ensemble of 500 independent realizations of the anisotropy landscape. Before applying the driving current, the system is relaxed to a local energy minimum. Consequently, the initial domain-wall position varies slightly among different disorder realizations due to interactions with the anisotropy landscape. To eliminate variability associated with the initial conditions, all domain-wall trajectories are shifted such that the initial wall position corresponds to $x(t=0)=0$. A transmission event is registered when the domain wall reaches the detector located at $x=400$~nm. Since thermal fluctuations are not included in the present model, any continued propagation after temporary pinning originates solely from the intrinsic micromagnetic dynamics of the extended domain wall. To distinguish transient slowing of the domain-wall motion from a genuinely pinned state, the simulation is continued until the domain wall reaches the detector or its position remains effectively unchanged ($<0.2$ nm) for $0.2$~ns. In the latter case, the wall is classified as pinned and the simulation is terminated.

\begin{figure*}
\centering
\includegraphics[width=5.9cm]{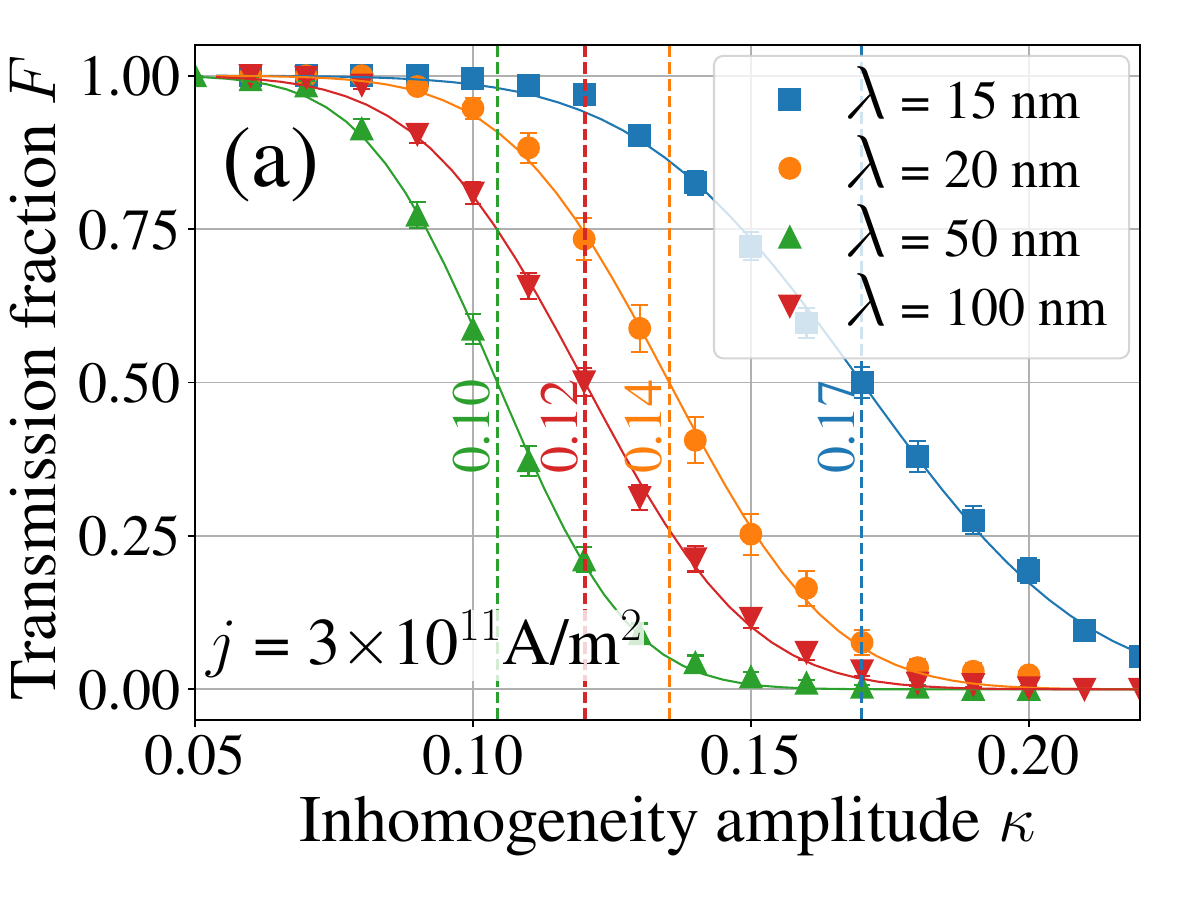}
\includegraphics[width=5.9cm]{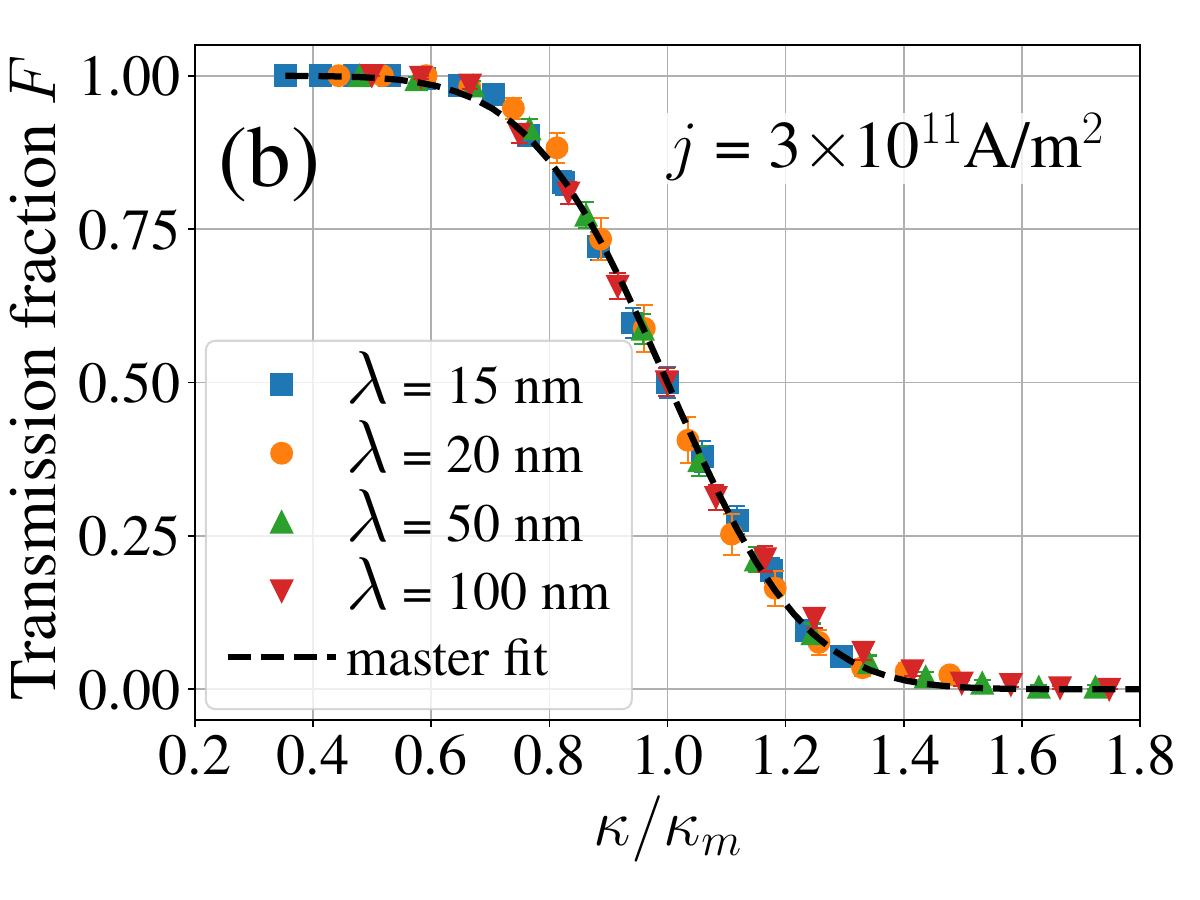}
\includegraphics[width=5.9cm]{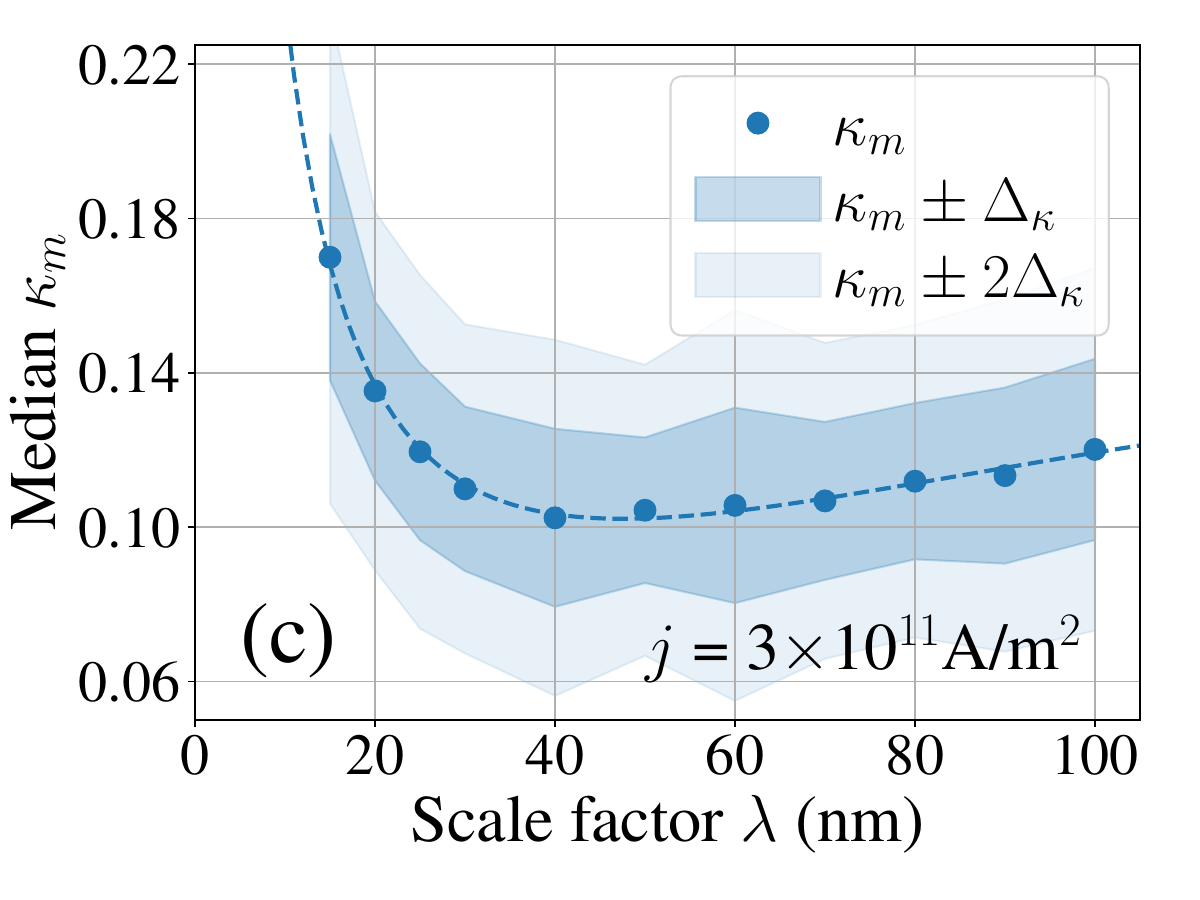}

\caption{\label{depin_lambda}
(a) Domain-wall transmission fraction $F$ as a function of anisotropy inhomogeneity amplitude $\kappa$  for various scale factors $\lambda$  at current density $j=3\times 10^{11}\, \mathrm{A/m^2}$. The probability is calculated for a racetrack of length 400 nm. Lines of identical color represent fits to the simulation data using the complementary Gaussian error-function $\bar {\Phi }(x)$. Vertical lines intersect the corresponding transmission curves at $F=0.5$ and indicate the median disorder amplitudes $\kappa_m$, whose values are labeled at the intersections. (b) Transmission fraction $F$  plotted versus the reduced amplitude $\kappa /\kappa_m$. (c) Median $\kappa_m$ as a function of scale factor $\lambda$. The shaded regions indicate the intervals $\kappa_m\pm\Delta_{\kappa}$ and $\kappa_m\pm2\Delta_{\kappa}$, where transmission fraction changes the most. The blue dashed line is a guide to the eye only.}
\end{figure*}

Since each realization of the anisotropy landscape produces a different sequence of local pinning barriers, the current density required for successful transmission becomes realization dependent. Repeating the simulations for many independent disorder realizations therefore allows us to determine the transmission fraction $F(j)$, defined as the fraction of domain walls that successfully reach the detector as a function of the applied current density $j$.
Figure~\ref{depin_prob}(a) shows the transmission fraction as a function of current density for several values of the anisotropy inhomogeneity amplitude $\kappa$ at a fixed scale factor $\lambda=25$~nm. The numerical data are accurately described by the Gaussian error-function profile
\begin{equation}
\Phi\left(j\right)=
\frac{1}{2}
\left[
1+
\mathrm{erf}
\left(
\frac{j-j_m}
{\sqrt{2}\Delta_j}
\right)
\right],
\label{Gaussian_fit}
\end{equation}
where $j_m$ denotes the median of the transmission curve ($\Phi=0.5$), while $\Delta_j$ is the transition width. 
The Gaussian error-function fits provide a convenient quantitative description of the transmission curves in terms of the median current density and the transition width.

As expected, increasing the disorder amplitude shifts the transmission curves toward larger current densities, indicating that stronger driving currents are required for successful transmission. At the same time, the transition region broadens, reflecting increased realization-to-realization variability.
To compare different disorder strengths, the current density is normalized by the median current density $j_m$. Remarkably, all transmission curves collapse onto a single master curve, as shown in Fig.~\ref{depin_prob}(b), where the width parameter $\Delta_j$ is about $0.18 j_{m}$. 
Similar scaling behavior and data collapse have previously been reported for stochastic domain-wall dynamics and creep motion in ferromagnetic nanowires \cite{Kim2009,Kim_2015}. Although the microscopic origin of stochasticity differs in the present system, the emergence of a master curve suggests common statistical features of disorder-controlled domain-wall transport.

The characteristic parameters extracted from the Gaussian error-function fits are summarized in Fig.~\ref{depin_prob}(c)--(d). The median current density $j_m$ increases approximately linearly with the disorder amplitude $\kappa$, indicating that stronger anisotropy fluctuations require larger driving currents for successful transmission. At the same time, the transition width $\Delta_j$ increases with the disorder amplitude $\kappa$ and thus also with the median current density $j_m$.
Figures~\ref{depin_prob}(e) and (f) present an equivalent inverse characterization of the transmission statistics. Instead of evaluating the transmission fraction for a given current density, we determine the current density $j_F$ required to achieve a prescribed transmission fraction using the inverse Gaussian error function, as detailed in Appendix~\ref{app:inverse_erfc}. Over the investigated parameter range, $j_F$ depends approximately linearly on the disorder amplitude $\kappa$ for all considered transmission fractions. Moreover, the slopes of these linear dependences exhibit an inverse Gaussian error-function dependence, indicating that the inverse representation preserves the same underlying behavior of the transmission curves.

Next, we investigate the influence of the characteristic scale factor $\lambda$ of the anisotropy landscape. Figure~\ref{depin_lambda}(a) shows the transmission fraction as a function of disorder amplitude $\kappa$ for several values of $\lambda$. Similar to the current-dependent transmission statistics discussed above, the numerical data follow a complementary Gaussian error-function dependence, as detailed in Appendix~\ref{app:complementary_erfc}. For each scale factor $\lambda$, we determine the median disorder amplitude $\kappa_m$ corresponding to a transmission fraction of $F=0.5$. Replotting the data as a function of the reduced disorder amplitude $\kappa/\kappa_m$ leads to a collapse of all curves onto a single master curve, as shown in Fig.~\ref{depin_lambda}(b) with a width parameter $\Delta_{\kappa} \approx 0.18 \kappa_m$.

The extracted values of $\kappa_m$ are shown in Fig.~\ref{depin_lambda}(c). The median disorder amplitude exhibits a shallow minimum at an intermediate scale factor, corresponding to $\lambda=40$~nm. At this scale, the correlation length of the anisotropy inhomogeneity is $\mathcal{L}=3.82$~nm (see Fig.~\ref{Fig_Corr_length}(a) in Appendix~\ref{App_Corr_length1}), which is close to the intrinsic domain-wall width $\delta\approx\sqrt{A_{\rm ex}/K_u^0}$, which for the micromagnetic parameters used here is equal to $3.16$~nm  \cite{Kronmuller2019}. This result suggests that transport suppression is most effective when the correlation length of the anisotropy landscape approaches the domain-wall width, consistent with previous studies of domain-wall pinning in weakly disordered ferromagnetic films \cite{Thiaville_2020_PhysRevResearch}. 

Overall, the results presented in Figs.~\ref{depin_prob} and \ref{depin_lambda} demonstrate that domain-wall transmission through structurally disordered racetracks exhibits robust Gaussian error-function scaling and data collapse. Despite the microscopic complexity of the anisotropy landscape, the transmission process can be quantitatively characterized by a small set of effective statistical parameters, including the median current density $j_m$, the median disorder amplitude $\kappa_m$, and the corresponding transition widths $\Delta_j$ and $\Delta_{\kappa}$. Together, these parameters provide a compact statistical description of both the average transport behavior and its stochastic fluctuations.

\section{Transport Statistics and Fluctuation Dynamics}

While the transmission statistics discussed in the previous section characterize the fraction of disorder realizations that result in successful domain-wall transport, they do not directly describe the variability of the propagation process. To obtain a more complete picture, we analyze the statistical fluctuations of domain-wall trajectories. This analysis characterizes how the quenched anisotropy disorder leads to realization-to-realization variations in the propagation dynamics.

For each realization of the anisotropy landscape, the domain-wall position $x(t)$ is extracted as a function of time. Statistical averages are evaluated over an ensemble of independent realizations of the disorder. Consequently, the observed trajectory fluctuations originate from realization-to-realization variations of the quenched anisotropy landscape.

To characterize the disorder-induced spread of domain-wall trajectories, we analyze the variance of the domain-wall position,
\begin{equation}
\sigma_\text{DW}^2(t)=
\left\langle
\left[
x(t)-\langle x(t)\rangle
\right]^2
\right\rangle,
\end{equation}
where the averaging is performed over independent realizations of the anisotropy landscape.

\begin{figure*}
\centering
\includegraphics[width=5.9cm]{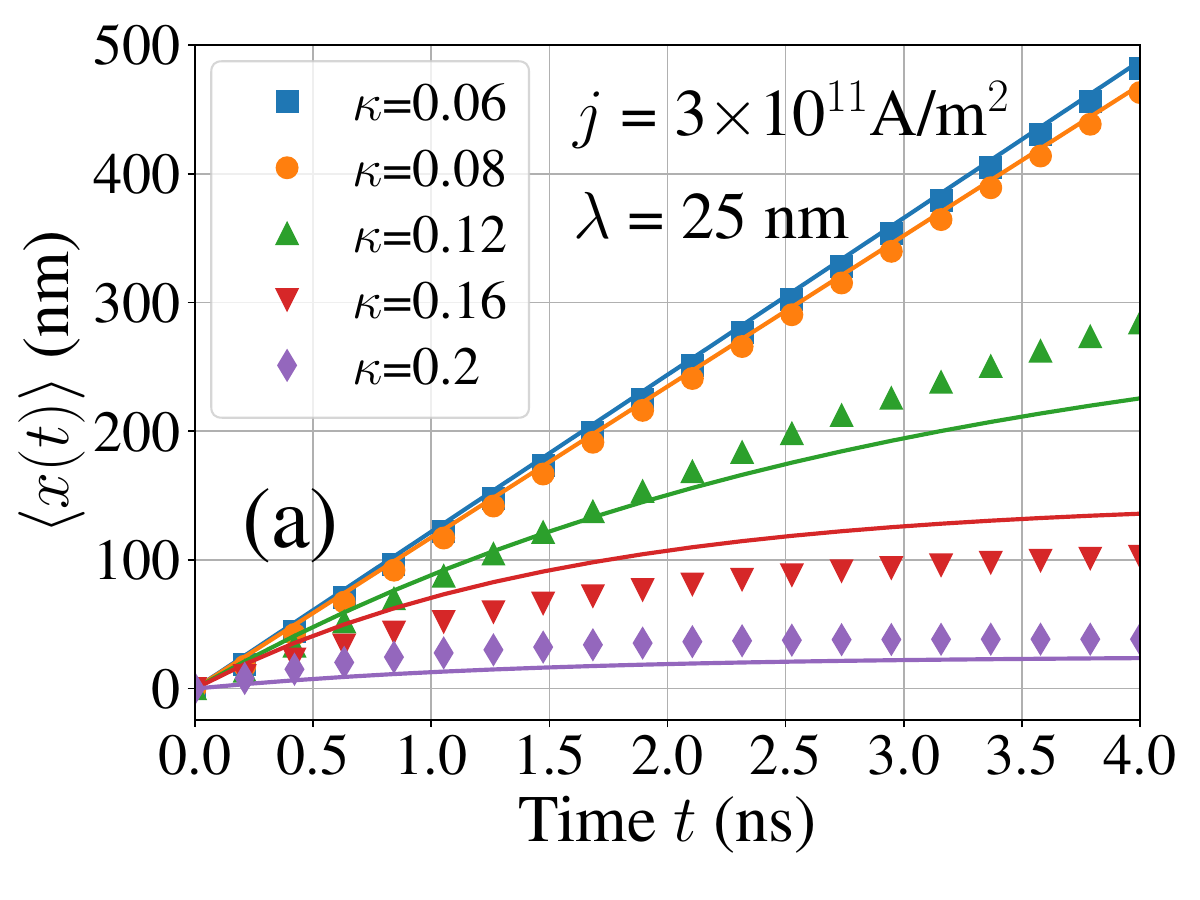}
\includegraphics[width=5.9cm]{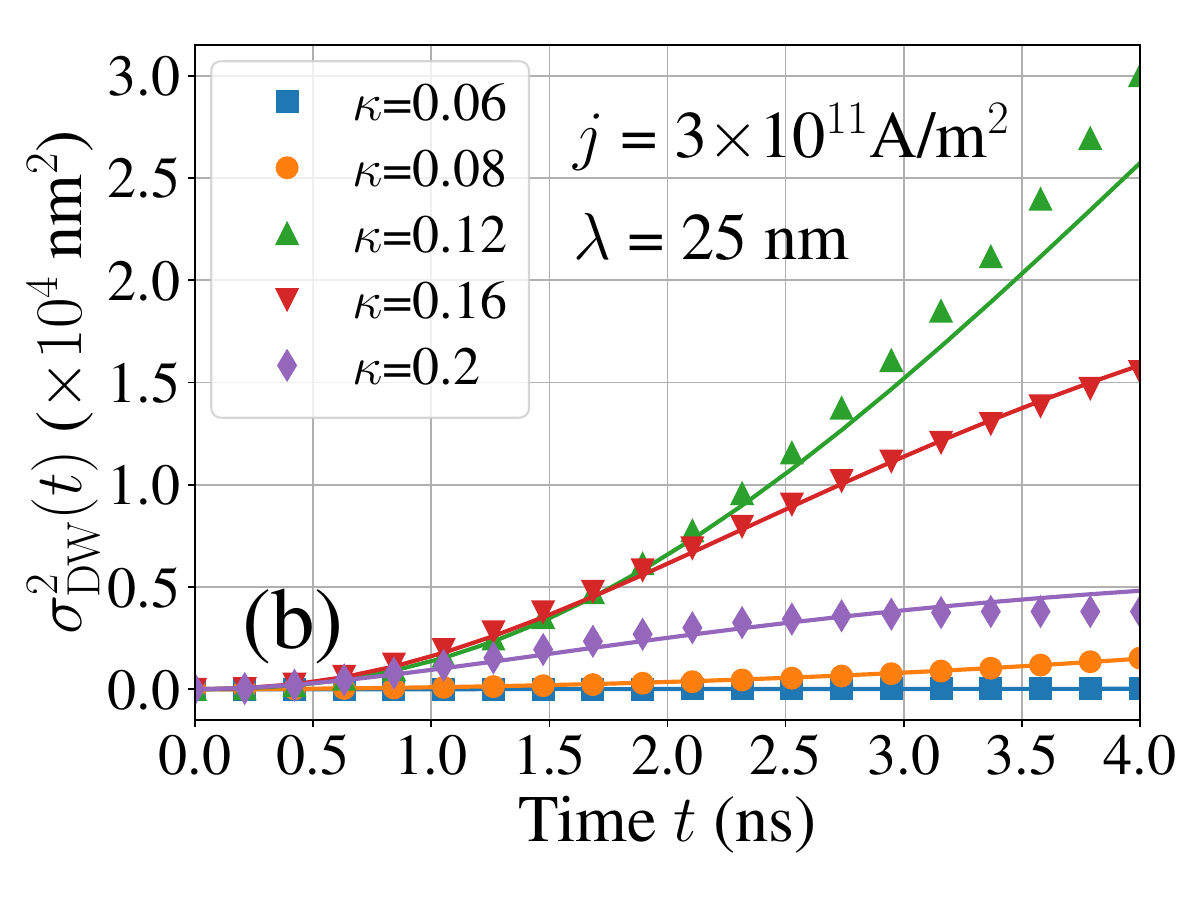}
\includegraphics[width=5.9cm]{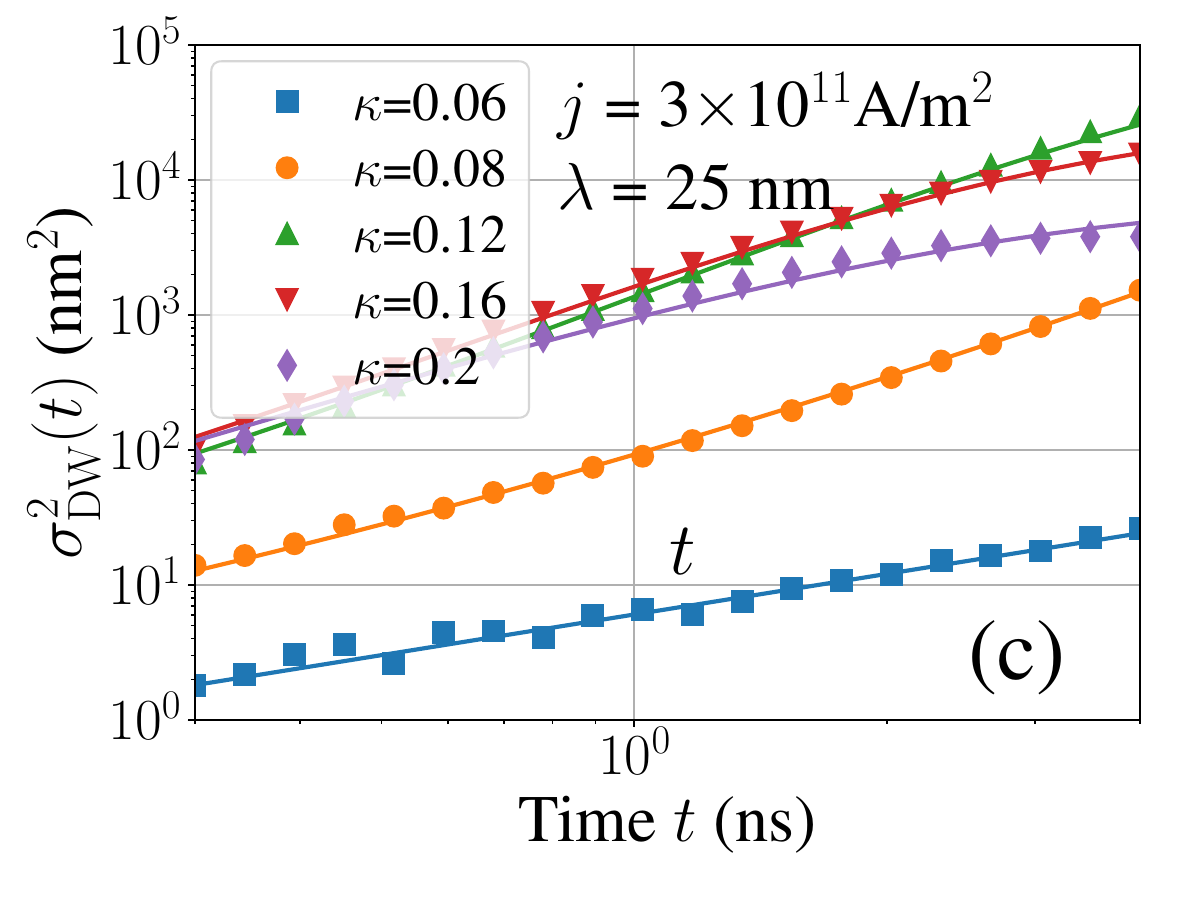}
\caption{\label{MSD}
(a) Ensemble-averaged domain-wall position $\langle x(t)\rangle$ as a function of time for different anisotropy inhomogeneity amplitudes $\kappa$ at fixed scale factor $\lambda=25$ nm and current density $j=3\times10^{11}$ A/m$^2$.
(b) Variance $\sigma_\text{DW}^2(t)$ of the domain-wall position calculated for the same set of parameters in linear-linear scale. (c) Variance $\sigma_\text{DW}^2(t)$ of the domain-wall position calculated for the same set of parameters in log-log scale. Solid lines represent the results from the analytical model (see details in the main text)}.
\end{figure*}

Figure~\ref{MSD}(a) shows the ensemble-averaged domain-wall position as a function of time for several values of the anisotropy inhomogeneity amplitude $\kappa$ at a fixed scale factor $\lambda=25$~nm and current density $j=3\times10^{11}$~A/m$^2$. For weak disorder, the domain wall propagates almost uniformly across the racetrack, resulting in an approximately linear increase of the mean position with time. As the disorder amplitude increases, the average propagation velocity gradually decreases due to stronger interactions with the inhomogeneous anisotropy landscape. Consequently, the mean propagation distance reached within the observation time is progressively reduced.

The corresponding trajectory variances are presented in
Fig.~\ref{MSD}(b), while Fig.~\ref{MSD}(c) shows the same data on a
log-log scale to better resolve their temporal evolution. For the
weakest disorder, $\kappa=0.06$, the variance increases approximately
linearly with time over the considered interval. This behavior is
consistent with the absence of pinning points in the corresponding
realizations, resulting in similar propagating trajectories. For
$\kappa=0.08$ and 0.12, the variance continues to increase throughout
the observation time window, indicating that the pinning-induced
saturation has not yet been reached. In this intermediate-disorder
regime, transmitted and non-transmitted trajectories coexist within
the ensemble, with some domain walls propagating over a substantial
fraction of the racetrack while others become trapped at relatively
short distances. This broad distribution of propagation distances
leads to an enhanced trajectory variance. For the stronger disorder,
$\kappa=0.16$ and 0.2, the variance approaches a plateau within the
considered time interval. In this regime, most domain walls become
pinned at relatively short distances, limiting the further growth of
the variance. Moreover, increasing the disorder amplitude from
$\kappa=0.16$ to 0.2 results in both a lower stationary variance and a
shorter saturation time.

To describe the observed time dependence of the DW position variance,
we introduce a simple statistical model in which the DW moves with an
average velocity $v$ until it becomes pinned at a random position $X$.
The corresponding trajectory is represented as
\begin{equation}
x(t)=vt\,\theta(X-vt)+X\,\theta(vt-X),
\label{app_eq_traj}
\end{equation}
where $\theta(x)$ denotes the Heaviside step function. 
Taking cue from micromagnetic simulation data (see Fig.~\ref{Fig_pinning_pdf}(a)) we consider the distribution of the pinning positions $p(X)$ represented by a localized peak combined with an exponential spread
\begin{equation}
p(X) = p_0 \delta(X) + (1 - p_0) \frac{1}{\Lambda} \exp\left(-\frac{X}{\Lambda}\right)
\label{eq_pdf_exponential}
\end{equation}
where $p_0$ is the probability of instantaneous pinning near the initial position, while $\Lambda$ defines the characteristic length scale. This choice of $p(X)$ leads to the DW mean position and variance given respectively as
\begin{equation}
    \left<x(t)\right>= (1-p_0)\Lambda\left(1-\exp\left(-\frac{vt}{\Lambda}\right)\right)
    \label{eq_mean_closed}
\end{equation}
and
\begin{eqnarray}
  \sigma_{\text{DW}}^2(t) = &(1 - p_0) \Lambda^2\left[ 1 + p_0 - 2\left(\frac{vt}{\Lambda} + p_0\right)\exp\left(-\frac{vt}{\Lambda}\right)\right. \nonumber\\
  &-\left.(1 - p_0) \exp\left(-\frac{2vt}{\Lambda}\right) \right].
\label{eq_var_closed}  
\end{eqnarray}
We refer to Appendix~~\ref{app:msd_model2} for details of the derivation. 

The simulation results exhibit an approximately linear
growth of the position variance before pinning becomes significant
(see Fig.~\ref{Fig_pinning_pdf}(b)). This is also seen as the jiggling motion of unpinned DWs, see Fig.~\ref{Displacements}(f). However, Eq.~(\ref{eq_var_closed}) leads to $\sigma_{\text{DW}}^2(t)\sim t^2$ at short times. In order to account for the observations from simulations,  we consider diffusion, leading to the following modified expression for the DW position variance
\begin{equation}
\begin{split}
\sigma_{\text{DW}}^2(t)
={}&(1-p_0)\Lambda^2
\left[
1+p_0-(1-p_0)
\exp\left(-\frac{2vt}{\Lambda}\right)
\right.\\
&\left.
-\left(
\frac{2vt}{\Lambda}
-\frac{Dt}{\Lambda^2}
+2p_0
\right)
\exp\left(-\frac{vt}{\Lambda}\right)
\right],
\end{split}
\label{eq_var_final}
\end{equation}
where $D$ is the coefficient characterizing the linear growth of the
position variance of the unpinned trajectories. 

Although the proposed model does not capture the full complexity of
stochastic DW dynamics, it provides a reasonable description of the
micromagnetic simulation results. The symbols in Fig.~\ref{MSD} correspond to the simulation data. In Fig.~\ref{MSD}(b,c) the solid lines correspond to Eq.~(\ref{eq_var_final}) with $\Lambda$ and $p_0$ treated as fitting parameters while $D$ and $v$ are obtained from micromagnetic simulation results. In particular, $D$ is obtained from the  $\log-\log$ plot of $\sigma_{\text{DW}}^2(t)$ vs. $t$ for unpinned trajectories, as shown in Fig.~\ref{Fig_pinning_pdf}(b). Mean DW velocities $v$ are obtained by linearly fitting $x(t)$ over the spatial interval spanning from $\lambda$ to the maximum attained position  (see Fig.~\ref{Displacements}(f) for representative trajectories). Choosing $\lambda$ as the lower bound ensures that the DW has traversed at least one characteristic period of the disorder landscape, effectively filtering out initial local relaxation and instant initial pinning.
The fitted parameters $\Lambda$ and $p_0$ from Fig.~\ref{MSD}(b,c) are then used in Eq.~(\ref{eq_mean_closed}) to plot the solid lines in Fig.~\ref{MSD}(a). The good agreement demonstrates that the
observed time dependence of the DW variance can be captured
by the combined effects of the stochastic fluctuations of unpinned
trajectories and the statistical distribution of pinning positions.

The characteristic time to reach the plateau on the variance curve is expected to increase as the disorder amplitude decreases and may therefore exceed the time window considered here. We focus on timescales up to 5~ns, corresponding to typical current-pulse durations in racetrack devices \cite{Guan2021,JaeChun_Science_2024} and encompassing the characteristic timescale of the present simulations. A systematic investigation of the temporal scaling at longer times is therefore beyond the scope of the present study. Extending the analysis to longer times may reveal a crossover from the initial growth of trajectory fluctuations to an intermediate-time regime and, eventually, to saturation associated with the increasing fraction of pinned domain walls.

A similar behavior was observed for domain-wall dynamics in racetracks with line-edge roughness \cite{hlushchenko_LER_2026}. Despite the different microscopic origins of the disorder, the trajectory variance in both systems exhibits a pronounced non-monotonic dependence on the disorder amplitude, together with a clear tendency toward saturation for strong disorder. This qualitative similarity suggests that the suppression of trajectory fluctuations at strong inhomogeneity may be a more general feature of disorder-driven domain-wall dynamics.


\section{Conclusions and Outlook}

We have investigated how spatially inhomogeneous magnetic anisotropy affects current-driven domain-wall transport in ferromagnetic racetracks. By representing structural disorder with two-dimensional fractal Weierstrass--Mandelbrot functions, we independently varied the strength and spatial scale of the anisotropy fluctuations. The simulations show that quenched structural disorder can by itself generate substantial realization-to-realization variability in domain-wall motion, even if the role of thermal fluctuations is neglected.


The transmission statistics reveal distinct effects of the amplitude and spatial scale of the anisotropy inhomogeneity. For a fixed disorder scale, the current density required for transmission increases linearly with the disorder amplitude. Moreover, the transmission curves for different disorder amplitudes collapse onto a common master curve when the current density is normalized by the median transmission current density, indicating a robust scaling of the current-dependent transmission response. The median disorder amplitudes corresponding to the transmission factor $0.5$ exhibit a non-monotonic dependence on the disorder scale, with the strongest suppression occurring when the correlation length of the anisotropy inhomogeneity is comparable to the intrinsic domain-wall width. These results demonstrate that the domain wall transport is governed not only by the strength of the disorder but also by the relation between its spatial correlation length and the characteristic width of the domain wall.

The ensemble statistics of domain-wall trajectories reveal a pronounced non-monotonic dependence of the trajectory variance on disorder amplitude. The variance is largest in the intermediate-disorder regime, where mobile and strongly trapped domain walls coexist and produce a broad distribution of propagation distances. At stronger disorder, enhanced pinning increasingly restricts domain-wall propagation, leading to a reduction and eventual saturation of the trajectory variance. These results are well described by the proposed phenomenological model.

Together, these results provide a quantitative basis for connecting
stochastic domain-wall transport with the properties of structural
disorder in magnetic racetracks. The ability to vary the amplitude and
spatial scale of the anisotropy landscape independently makes it
possible to distinguish their respective contributions to transport
variability and reliability. Future work could extend this approach by
systematically varying the domain-wall width and racetrack geometry,
thereby clarifying how the intrinsic length scales of the magnetic
texture and device interact with those of the disorder landscape.
An important next step will be to investigate the combined influence of
thermal fluctuations and structural disorder within the same racetrack
device. Their interplay may be particularly important for determining
the stochastic transport characteristics and the practical operating
window of racetrack technologies  \cite{Jeon2026Opportunities}. This framework could further be
extended to spatial inhomogeneities of other micromagnetic parameters,
such as the Dzyaloshinskii--Moriya interaction, to assess their combined
effects on domain-wall transport \cite{Kawaguchi_npj_2021}.
Extending the analysis to other magnetic textures, including skyrmions,
could further reveal how structural disorder shapes stochastic
propagation and transport statistics across different types of
topological magnetic textures. Such studies may help establish broader
design principles for controlling disorder-induced reliability limits
in racetrack-based spintronic devices.

Beyond their significance for understanding disorder-controlled transport in racetrack devices, the present results establish a physical mechanism for probabilistic domain-wall dynamics with potential relevance to probabilistic spintronic computing \cite{Camsari2017,Camsari2019,Finocchio2024_Roadmap}. The observed sigmoidal transmission characteristics provide a key ingredient for probabilistic information processing. At the same time, the device-to-device variability originating from different realizations of the anisotropy landscape naturally suggests potential applications, for example, in physically unclonable functions (PUFs) \cite{Zhang2020,Gao2020,Ghosh2016}. More broadly, the rich dynamics of propagating domain walls, including stochastic transmission, propagation, and arrival-time statistics, may provide new opportunities for probabilistic and event-driven spintronic computing architectures.

\section*{Acknowledgments}
\label{sec:ack}
KIP, AMS, and AVC acknowledge support from the BMFTR project 01DK24006 PLASMA-SPIN-ENERGY. AVH, OLA, and MIB acknowledge support from the National Research Foundation of Ukraine, Project No. 2025.07/0086 (Excellent Science in Ukraine 2026--2028).

\appendix

\section{Properties of the W--M Function}

In this section, we evaluate the variance of the surface height and correlation length of the W-M function.

\subsection{Variance of W-M function}\label{App_RMS}
We evaluate the variance $\sigma^2_\mathcal{W}$ of the W–M function $\mathcal{W}$ given by Eq. (\ref{W_M}),
\begin{equation}
\begin{split}\label{RMS_A}
&\sigma^2_\mathcal{W}={\langle\mathcal{W}^2(x,y)\rangle}.
\end{split}
\end{equation}
Next, we rewrite the W-M functions as
\begin{equation}
\begin{split}\label{W_simp}
&\mathcal{W}(x,y)=\sum_{n,m}a_{m,n}X_{m,n}, \\
&a_{n}= \gamma_\mathcal{W}^{-n H}, \\
&X_{m,n}=\cos{\left( \omega_{m,n}-\phi(m,n)\right)}, \\
&\omega_{m,n}=k_n\left(x\cos{\theta_m} + y\sin{\theta_m}\right).
\end{split}
\end{equation}
The correlation function is given by
\begin{equation}
\begin{split}\label{W_2}
&\langle\mathcal{W}(x,y)^2\rangle=\sum_{n,m}\sum_{n',m'}a_{n}a_{n'}\langle X_{m,n}X_{m',n'}\rangle.
\end{split}
\end{equation}
Since the phases $\phi(m,n)$ are independent random functions, therefore for $(m,n)\neq(m',n')$ we obtain
\begin{equation}
\begin{split}\label{X_2}
&\langle X_{m,n}X_{m',n'}\rangle=\langle X_{m,n}\rangle\langle X_{m',n'}\rangle=0.
\end{split}
\end{equation}
Thus, the final form of $\sigma^2_\mathcal{W}$ reads as
\begin{equation}
\begin{split}\label{aver_W}
&\langle\mathcal{W}(x,y)^2\rangle=\sigma^2_\mathcal{W}=\sum_{n,m}a_{m}^2\langle X_{m,n}^2\rangle= \\ 
&\sum_{n,m}a_{m}^2\langle \cos^2{\left( \omega_{m,n}-\phi(m,n)\right)}\rangle=\frac{1}{2}\sum_{n,m}a_{n}^2= \\
&\frac{1}{2}\sum_{n=0}^{N-1}\sum_{m=1}^M\gamma_\mathcal{W}^{-2n (3-D)}=\frac{M}{2}\sum_{n=0}^{N-1}\gamma_\mathcal{W}^{-2n (3-D)}= \\ 
&{\frac{M}{2}\frac{1-\gamma_\mathcal{W}^{-2(3-D)N}}{1-\gamma_\mathcal{W}^{-2(3-D)}}}.
\end{split}
\end{equation}

\subsection{Correlation length}\label{App_Corr_length1}

\begin{figure*}
\centering
\includegraphics[width=5.9cm]{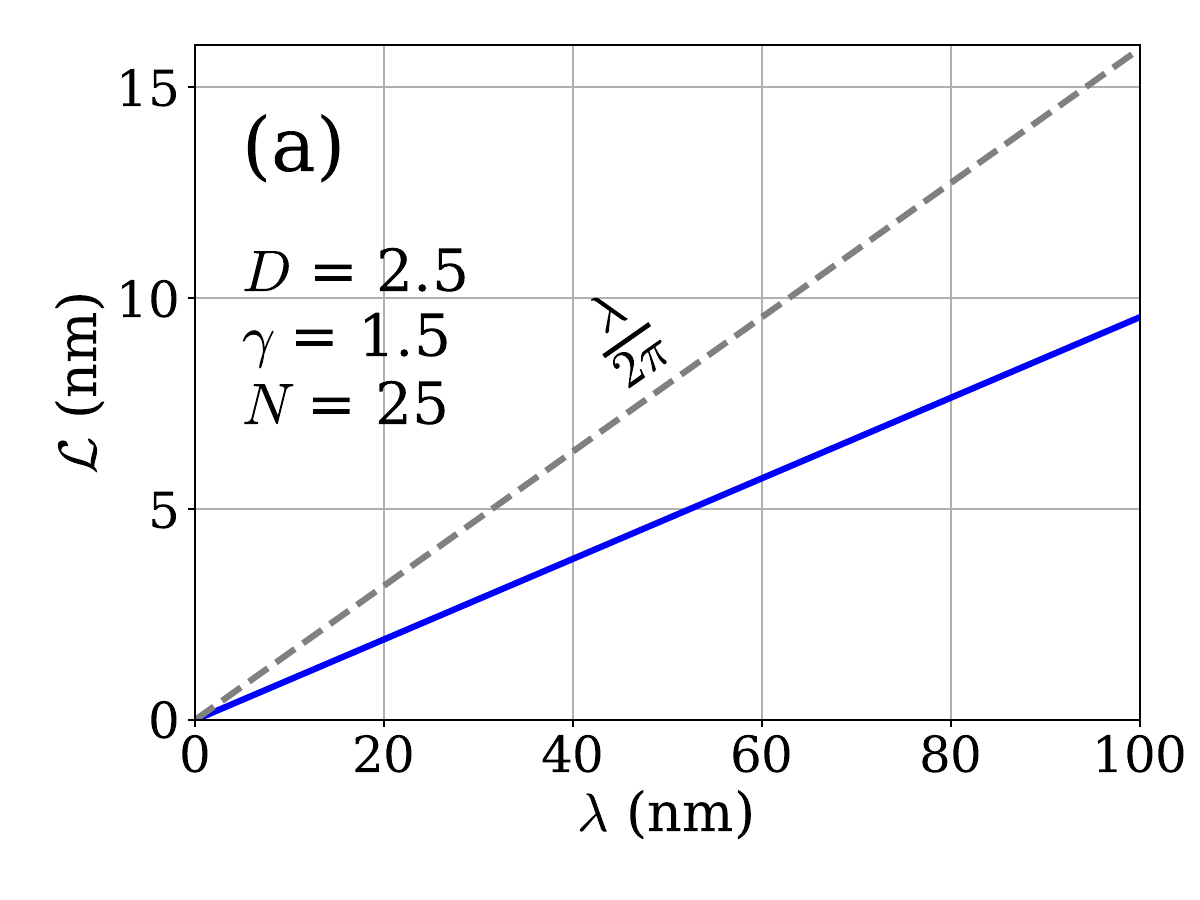}
\includegraphics[width=5.9cm]{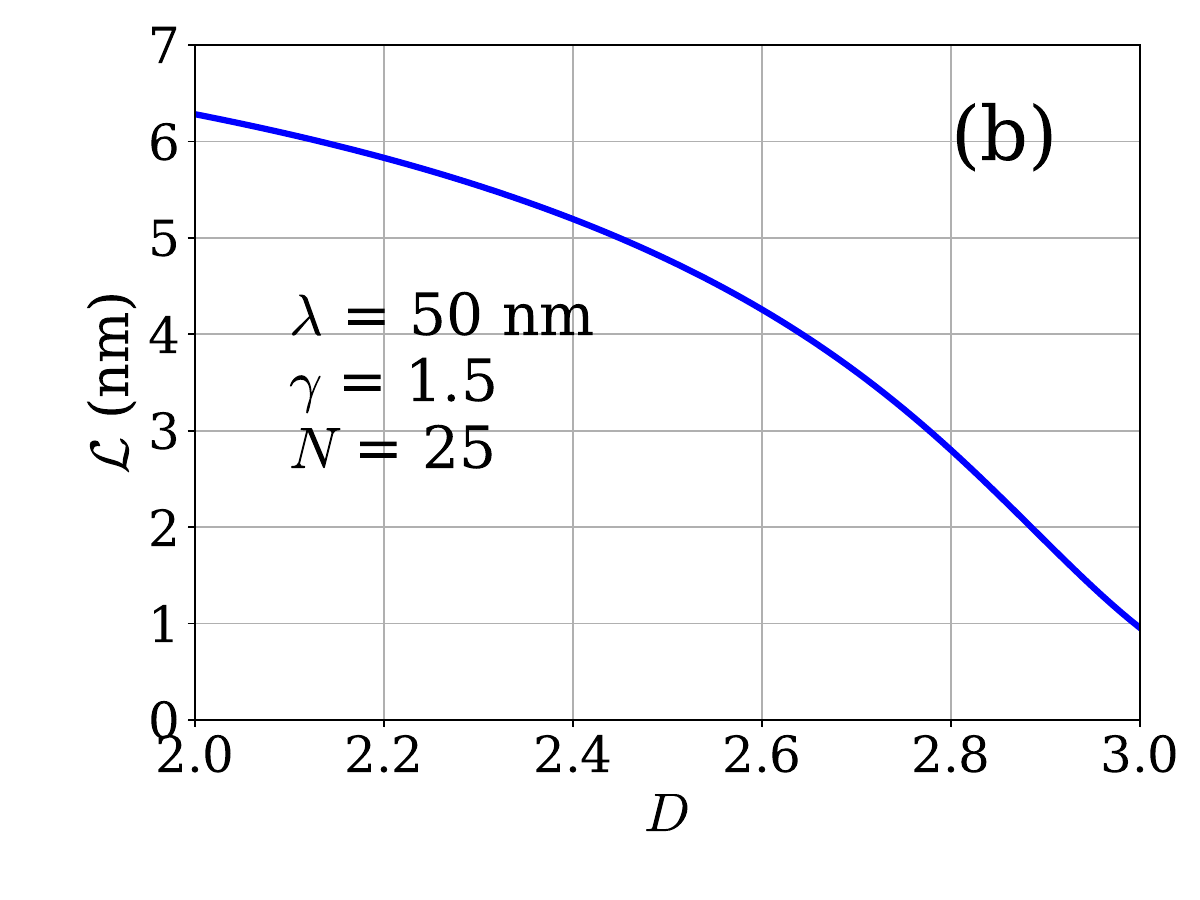}
\includegraphics[width=5.9cm]{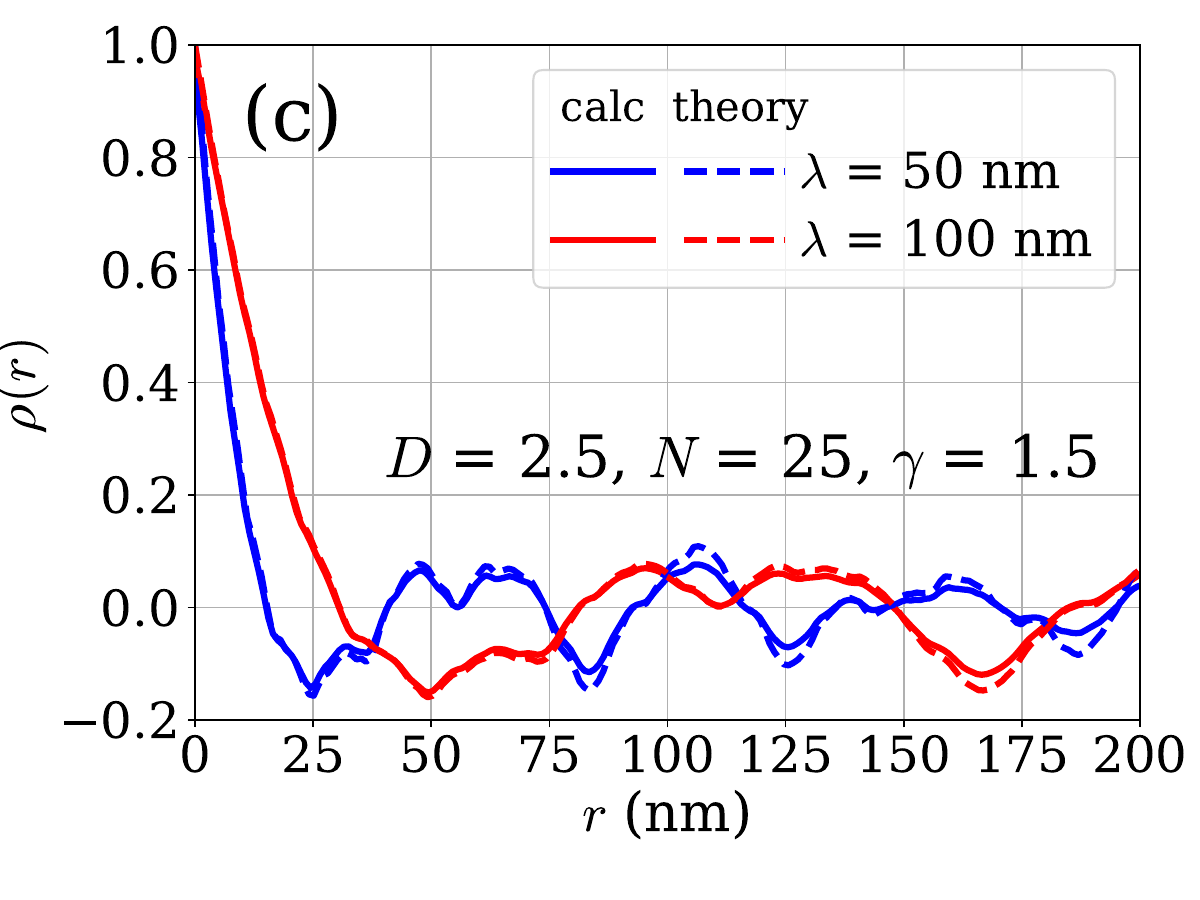}

\caption{\label{Fig_Corr_length}
Correlation length $\mathcal{L}$ and autocorrelation function $\rho(r)$ of the two-dimensional Weierstrass--Mandelbrot disorder landscape.
(a) Dependence of the correlation length $\mathcal{L}$ on the scale factor $\lambda$ for $D=2.5$, $\gamma_\mathcal{W}=1.5$, and $N=25$. The dashed line corresponds to the asymptotic relation $\mathcal{L}=\lambda/(2\pi)$. The horizontal dashed line marks the correlation length $\mathcal{L}=3.82$~nm corresponding to $\lambda=40$~nm, the scale factor at which the minimum of $\kappa_m$ is observed.
(b) Correlation length as a function of the fractal dimension $D$ for $\lambda=50$ nm, $\gamma_\mathcal{W}=1.5$, and $N=25$.
(c) Comparison of the normalized correlation function obtained numerically from disorder realizations (calc) and predicted by Eq.~(\ref{C_r_final}) (theory) for $\lambda=50$ nm and $100$ nm with $D=2.5$, $\gamma_\mathcal{W}=1.5$, and $N=25$.
}
\end{figure*}

To calculate the correlation length we use the definition of correlation function
\begin{equation}\label{corr_func}
\mathcal{C}(\textbf{r})=\langle\mathcal{W}(\textbf{r}_0)\mathcal{W}(\textbf{r}_0+\textbf{r})\rangle.
\end{equation}
Given the properties of the W-M function Eq. (\ref{W_M}), we can choose an arbitrary starting point $\textbf{r}_0$ and direction $\textbf{r}$, namely $\textbf{r}_0=\{0,0\}$ and $\textbf{r}=\{r,0\}$. 

W-M functions $\mathcal{W}(0)$ and $\mathcal{W}(r)$ take the form
\begin{equation}
\begin{split}\label{WM_corr}
&\mathcal{W}(0)=a_W\sum_{n=0}^{N-1} 
\sum_{m=1}^{M}\gamma_\mathcal{W}^{-n (3-D)}\cos{\left( \phi_{m,n}\right)}, \\
&\mathcal{W}(r)=a_W\sum_{n=0}^{N-1} 
\sum_{m=1}^{M}\gamma_\mathcal{W}^{-n (3-D)}\cos{\left( k_nr\cos{\theta_m}-\phi_{m,n}\right)}.
\end{split}
\end{equation}
To calculate the correlation function $\mathcal{C}(r)$, it is necessary to consider the product, which gives four sums according to $(m,n)$ and $(m',n')$. From the definition of the W-M function (see Eq. (\ref{W_M})) it follows that $\phi(m,n)$ are independent random variables uniformly distributed in the range $[0,2\pi)$. As a result, all cross terms disappear, and only terms with $m=m'$ and $n=n'$ remain. 
The correlation functions $\mathcal{C}(0)$ takes the form
\begin{equation}
\begin{split}\label{C_0}
\mathcal{C}(0)=&a^2_W\sum_{n=0}^{N-1} 
\sum_{m=1}^{M}\gamma_\mathcal{W}^{-2n (3-D)}\langle\cos^2{\left( \phi_{m,n}\right)}\rangle= \\
&\frac{a^2_W}{2}\sum_{n=0}^{N-1}\sum_{m=1}^{M}\gamma_\mathcal{W}^{-2n (3-D)},
\end{split}
\end{equation}
where we use
\begin{equation}
\begin{split}\label{cos_1}
\langle\cos^2{\left( \phi_{m,n}\right)}\rangle=\frac{1}{2}\langle(1+\cos{\left( 2\phi_{m,n}\right)}\rangle=\frac{1}{2}.
\end{split}
\end{equation}
The correlation functions $\mathcal{C}(r)$ takes the form
\begin{equation}
\begin{split}\label{C_r}
&\mathcal{C}(r)=a^2_W\sum_{n=0}^{N-1} 
\sum_{m=1}^{M}\gamma_\mathcal{W}^{-2n (3-D)}\times \\
&\langle\cos{\left( \phi_{m,n}\right)}\cos{(k_nr\cos{\theta_m}-\phi_{m,n})}\rangle=\\
&\frac{a^2_W}{2}\sum_{n=0}^{N-1}\sum_{m=1}^{M}\gamma_\mathcal{W}^{-2n (3-D)}\cos{(k_nr\cos{\theta_m})},
\end{split}
\end{equation}
where
\begin{equation}
\begin{split}\label{cos_2}
&\langle\cos{\left( \phi_{m,n}\right)}\cos{(k_nr\cos{\theta_m}-\phi_{m,n})}\rangle=\\
&\frac{1}{2}\langle\cos{\left(k_nr\cos{\theta_m}\right)}+\cos{(k_nr\cos{\theta_m}-2\phi_{m,n})}\rangle=\\
&\frac{1}{2}\cos{\left(k_nr\cos{\theta_m}\right)}.
\end{split}
\end{equation}
Using Eqs. (\ref{C_0})-(\ref{cos_2}) the normalized correlation function $\rho({r})=\mathcal{C}(r)/\mathcal{C}(0)$ reads
\begin{equation}
\begin{split}\label{norm_final}
&\rho({r})=\frac{\sum_{n=0}^{N-1}\sum_{m=1}^{M}\gamma_\mathcal{W}^{-2n (3-D)}\cos{(k_nr\cos{\theta_m})}}{M\sum_{n=0}^{N-1}\gamma_\mathcal{W}^{-2n (3-D)}},\\
&k_n=\frac{2\pi\gamma_\mathcal{W}^n}{\lambda}, \theta_m=\frac{2\pi m}{M}.
\end{split}
\end{equation}
Equation (\ref{norm_final}) contains a Riemann sum $\sum_{m=1}^{M}\cos{(k_nr\cos{\theta_m})}$, $\theta_m=\frac{2\pi m}{M}$, which is an approximation (for large $M$) of the following integral
\begin{equation}
\begin{split}\label{int}
&\sum_{m=1}^{M}\cos{(k_nr\cos{\theta_m})}\approx \\
&\frac{M}{2\pi}\int_0^{2\pi}{\cos{(k_nr\cos{\theta})}d\theta}=
MJ_0(k_nr).
\end{split}
\end{equation}
Therefore
\begin{equation}
\begin{split}\label{C_r_final}
&\rho(r)=\frac{\sum_{n=0}^{N-1}\gamma_\mathcal{W}^{-2n(3-D)}J_0(k_nr)}{\sum_{n=0}^{N-1}\gamma_\mathcal{W}^{-2n (3-D)}}
\end{split}
\end{equation}
By definition, the correlation length $\mathcal{L}$ is equal to
\begin{equation}
\begin{split}\label{cor_length_def}
&\mathcal{L}=\int_0^\infty \rho(r)dr=\\
&\frac{\sum_{n=0}^{N-1}\gamma_\mathcal{W}^{-2n(3-D)}\int_0^\infty J_0(k_nr)dr}{\sum_{n=0}^{N-1}\gamma_\mathcal{W}^{-2n(3-D)}}=\\
&\frac{\sum_{n=0}^{N-1}\gamma_\mathcal{W}^{-2n(3-D)}(1/k_n)}{\sum_{n=0}^{N-1}\gamma_\mathcal{W}^{-2n(3-D)}}=\\
&\frac{\lambda}{2\pi}\frac{\sum_{n=0}^{N-1}\gamma_\mathcal{W}^{-n(2(3-D)+1)}}{\sum_{n=0}^{N-1}\gamma_\mathcal{W}^{-n2(3-D)}}.
\end{split}
\end{equation}
Next, we use the properties of geometric progression for finite sums which results in 
\begin{equation}
\begin{split}\label{cor_length_def}
&\mathcal{L}=\frac{\lambda}{2\pi}\frac{(1-\gamma_\mathcal{W}^{-N(7-2D)})(1-\gamma_\mathcal{W}^{-2(3-D)})}{(1-\gamma_\mathcal{W}a^{-(7-2D)})(1-\gamma_\mathcal{W}^{-2N(3-D)})}.
\end{split}
\end{equation}
As follows from Eq.~(\ref{cor_length_def}), the correlation length $\mathcal{L}$ is proportional to the scale factor $\lambda$ of the Weierstrass--Mandelbrot random function $\mathcal{W}(x,y)$ (see also Fig.~\ref{Fig_Corr_length}(a)). Therefore, throughout the main text we use the scale factor $\lambda$ as the primary parameter characterizing the spatial scale of the disorder. The correlation length also depends on the fractal dimension $D$ and decreases with increasing $D$ for fixed $\lambda$, as illustrated in Fig.~\ref{Fig_Corr_length}(b). The normalized correlation functions obtained numerically from disorder realizations demonstrate a good agreement with those computed analytically with Eq.~(\ref{C_r_final}), see Fig.~\ref{Fig_Corr_length}(c).


\section{Fit Error-Function Representation}
\label{app:erfc}

\subsection{Inverse Gaussian Error-Function}
\label{app:inverse_erfc}

The inverse Gaussian error function $\Phi^{-1}(F)$, defined such that
$\Phi\!\left(\Phi^{-1}(F)\right)=F$, is used to determine the
corresponding current density for a prescribed transmission fraction
$F$,
\begin{equation}
j_F
=
j_m+\Delta_j\Phi^{-1}(F)
=
j_m+
\sqrt{2}\Delta_j\,
\mathrm{erf}^{-1}(2F-1).
\label{app_inverse_j}
\end{equation}
Here, $j_F$ denotes the current density required to achieve the
prescribed transmission fraction $F$.

The dependence of the fitted slope $\alpha(F)$ on the transmission
fraction can likewise be described by the inverse Gaussian error
function,
\begin{equation}
\alpha(F)
\approx
\bar{\alpha}
\left[
1+\Delta_\alpha\Phi^{-1}(F)
\right],
\label{app_inverse_alpha}
\end{equation}
where $\bar{\alpha}$ and $\Delta_\alpha$ are fitting parameters.

\subsection{Complementary Error-Function}
\label{app:complementary_erfc}

For the dependence of the transmission fraction on the disorder amplitude $\kappa$, the complementary form of the Gaussian error-function profile $\bar{\Phi}(x)=1-\Phi(x)$ is used, that is
\begin{equation}
\bar{\Phi}\left(
\kappa
\right)
=
\frac{1}{2}
\left[
1-
\mathrm{erf}
\left(
\frac{\kappa-\kappa_m}
{\sqrt{2}\Delta_\kappa}
\right)
\right].
\label{app_gaussian_kappa}
\end{equation}
Here, $\kappa_m$ denotes the median disorder amplitude, while $\Delta_\kappa$ characterizes the width of the Gaussian error-function.

\section {Statistical Model of Domain Wall Trajectory and Variance}
\label{app:msd_model2}
\begin{figure*}
\centering
\includegraphics[width=5.9cm]{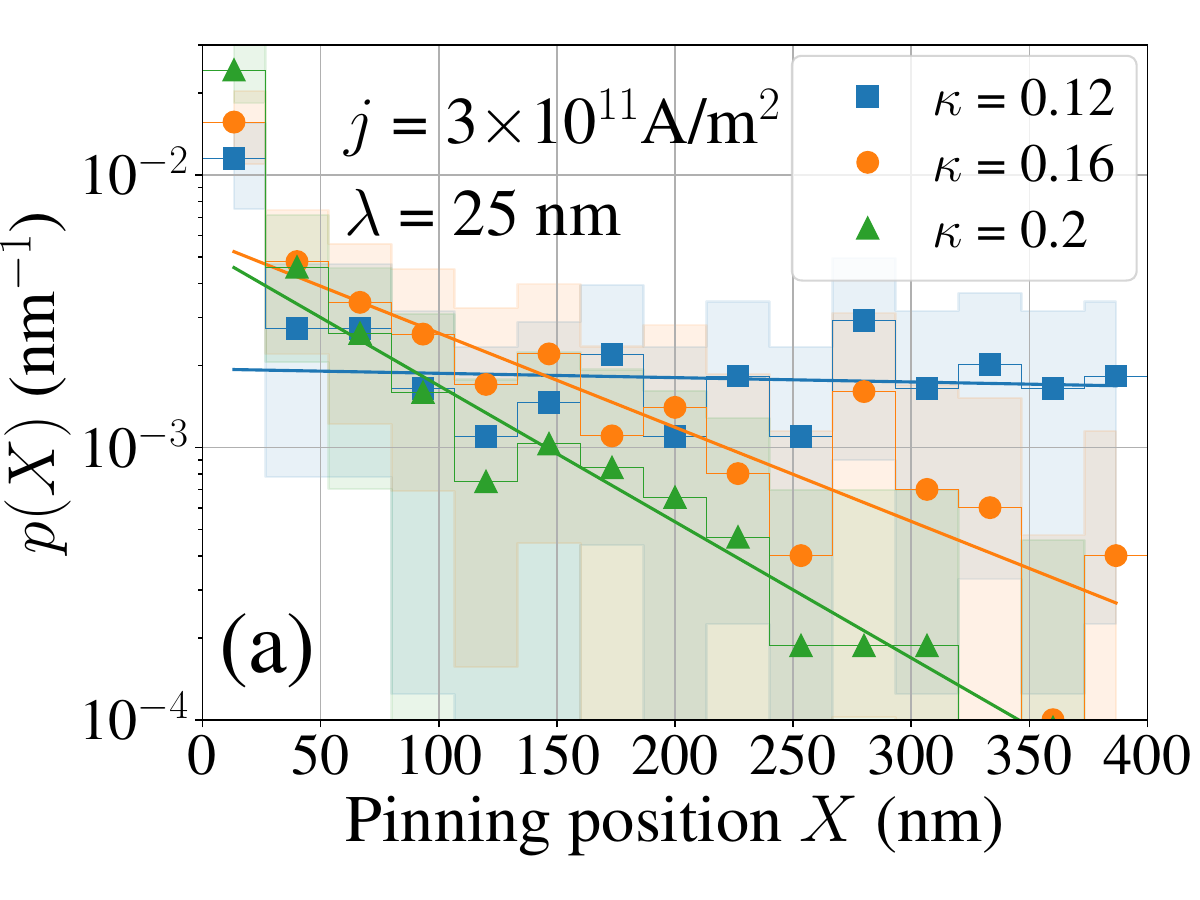}
\includegraphics[width=5.9cm]{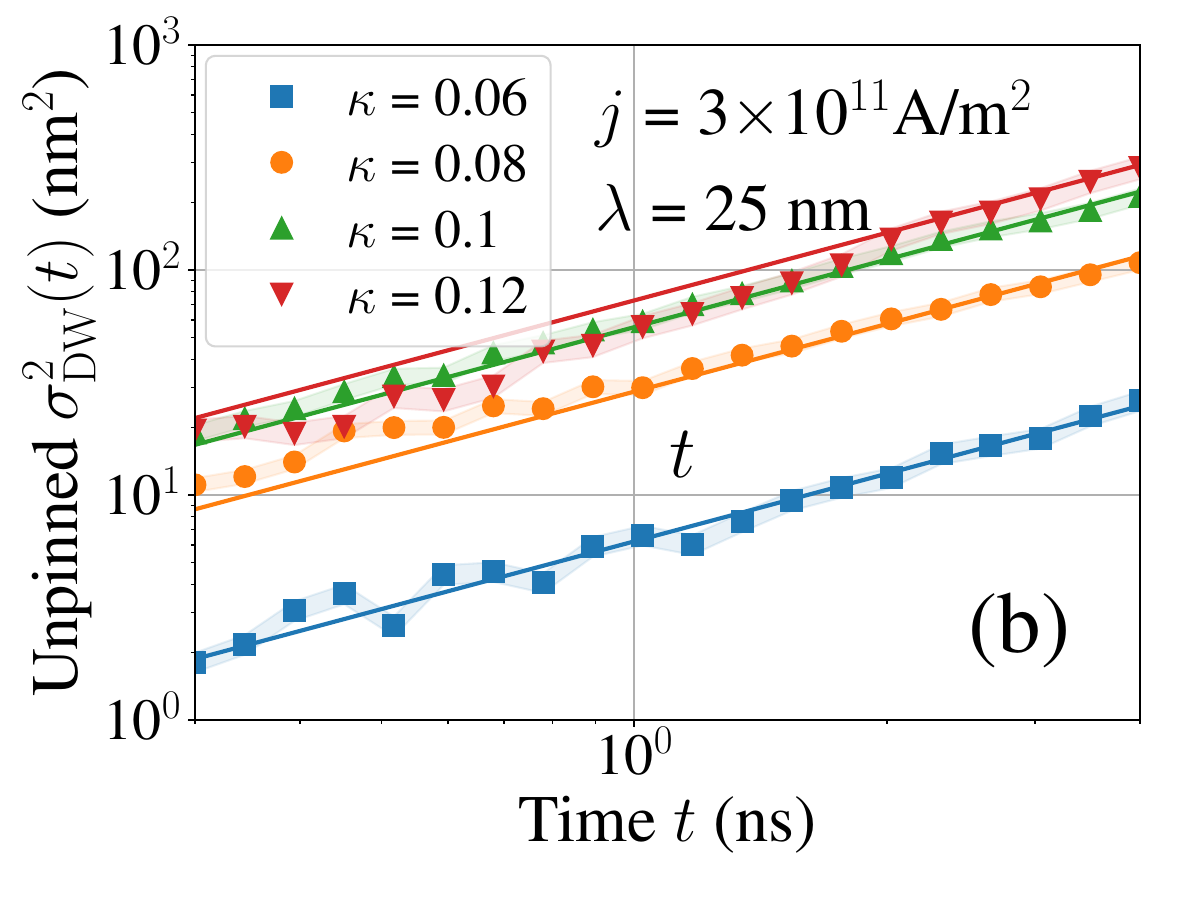}
\includegraphics[width=5.9cm]{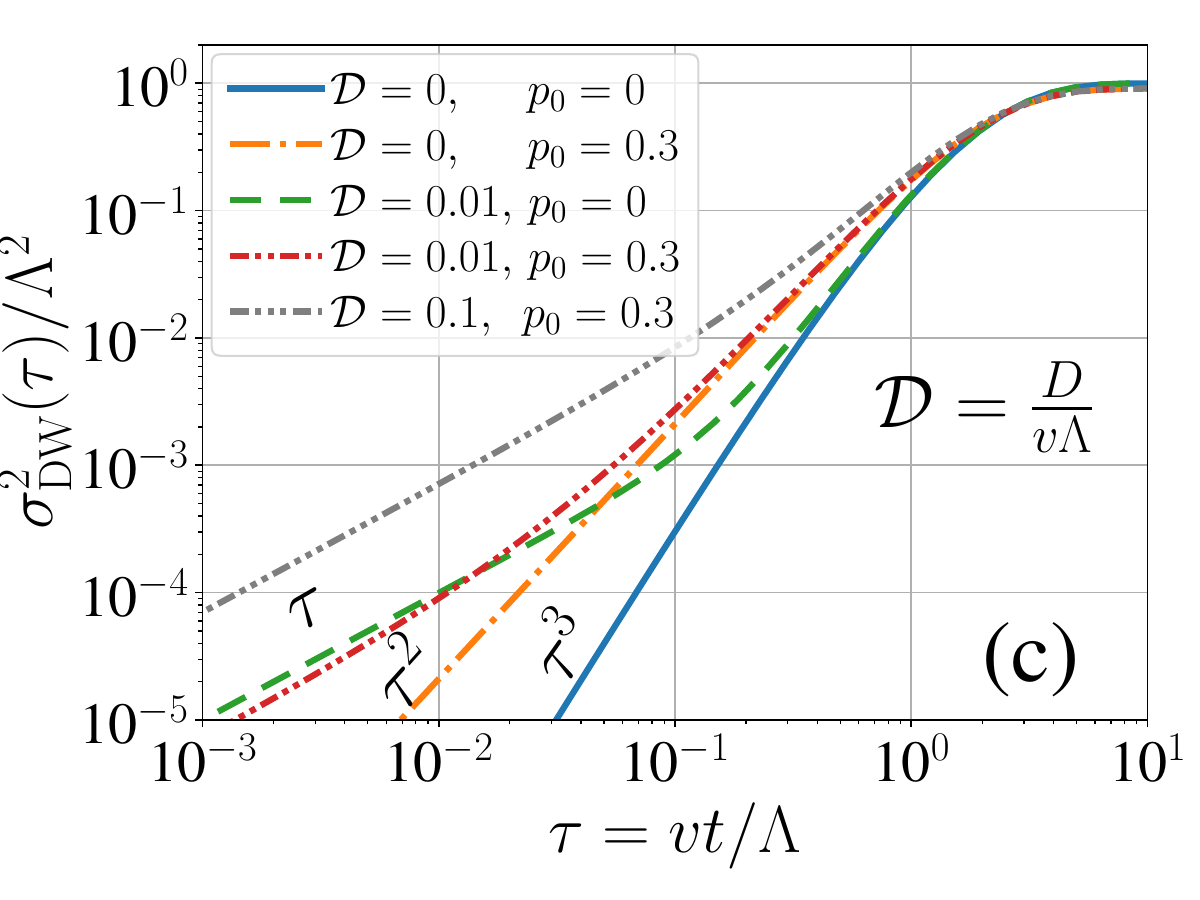}
\caption{\label{Fig_pinning_pdf} (a) The estimation of $p\left(X\right)$ from micromagnetic simulation data for $j=3\times10^{11}\mathrm{A/m}^2$ and $\lambda=25$~nm. Lines show exponential fits of the distribution tail. (b) The estimated DW position variance for unpinned trajectories only. Solid lines show linear approximation of the data. The uncertainties are within the symbol size. (c) The analytical time dependence of reduced DW position variance $\sigma_{\text{DW}}^2/\Lambda^2$ for various values of dimensionless parameters $\mathcal{D}$ and $p_0$. The shaded areas in (a) and (b) show the uncertainty estimates.
}
\end{figure*}

To approximate the domain wall (DW) dynamics under pinning effects, the trajectory $x(t)$ as a function of time $t$ can be treated as a combination of deterministic motion with mean velocity $v$ prior to pinning and a permanent stationary position $X$ following a pinning event
\begin{equation}
x(t) = vt \cdot \theta(X - vt) + X \cdot \theta(vt - X)
\label{app_eq_traj}
\end{equation}
where $\theta(x)$ denotes the Heaviside step function. The spatial pinning positions $X$ are random variables governed by a probability density function (PDF) $p(X)$. The probability $Q(t)$ that a DW remains unpinned up to time $t$ is expressed as
\begin{equation}
Q(t) = \int_{vt}^{\infty} p(X) dX
\label{app_eq_survival}
\end{equation}
So, the $n$-th moment of the DW position $x(t)$ is given by
\begin{equation}
\langle x^n(t) \rangle = (vt)^n \cdot Q(t) + \int_{0}^{vt} X^n p(X) dX
\label{app_eq_general_traj_moms}
\end{equation}
Let us introduce the time-dependent partial pinning moments
\begin{equation}
\left<X^n\right>_t = \frac{\int_{0}^{vt} X^n p(X) dX}{\int_{0}^{vt} p(X) dX} = \frac{\int_{0}^{vt} X^n p(X) dX}{1 - Q(t)}.
\label{app_eq_nth_pin_moms}
\end{equation}
Using Eqs.~(\ref{app_eq_general_traj_moms}) and (\ref{app_eq_nth_pin_moms}),  we obtain the DW position variance $\sigma_{\text{DW}}^2(t) = \left<x^2(t)\right> - \left<x(t)\right>^2$ in the next form:
\begin{equation}
\sigma_{\text{DW}}^2(t) = (1 - Q(t))\left[ \left<\delta X^2\right>_t + Q(t)\left( vt - \left<X\right>_t \right)^2\right]
\label{app_eq_var_traj}
\end{equation}
where $\left<\delta X^2\right>_t = \left<X^2\right>_t - \left<X\right>_t^2$.

The micromagnetic simulation data (see Fig.~\ref{Fig_pinning_pdf}(a)) indicate that the pinning distribution $p(X)$ can be represented by a localized peak combined with an exponential spread
\begin{equation}
p(X) = p_0 \delta(X) + (1 - p_0) \frac{1}{\Lambda} \exp\left(-\frac{X}{\Lambda}\right),
\label{app_eq_pdf_exponential}
\end{equation}
where $p_0$ is the probability of instantaneous pinning near the initial position, while $\Lambda$ defines the characteristic length scale. Evaluating the integrals in Eqs.~(\ref{app_eq_survival}-\ref{app_eq_nth_pin_moms}) we obtain 
\begin{equation}
\begin{split}
Q(t) &= (1 - p_0) e^{-\tau} \\
\frac{\left<X\right>_t}{\Lambda} &= (1 - p_0) \frac{1 - (1 + \tau) e^{-\tau}}{1 - (1 - p_0) e^{-\tau}} \\
\frac{\left<X^2\right>_t}{\Lambda^2} &= (1 - p_0) \frac{2 - \left[1 + (1 + \tau)^2\right] e^{-\tau}}{1 - (1 - p_0) e^{-\tau}},
\end{split}
\label{app_eq_exp_stats}
\end{equation}
where we introduced the dimensionless parameter $\tau = vt/\Lambda$.

Direct substitution of Eq.~(\ref{app_eq_exp_stats}) into the variance equation Eq.~(\ref{app_eq_var_traj}) leads to a closed-form solution for the normalized spatial variance:
\begin{equation}
\frac{\sigma_{\text{DW}}^2(t)}{\Lambda^2} = (1 - p_0) \left[ 1 + p_0 - 2(\tau + p_0) e^{-\tau} - (1 - p_0) e^{-2\tau} \right]
\label{app_eq_var_closed}
\end{equation}


The approximation of deterministic motion before pinning can be extended by accounting for the DW position variance before pinning. For unpinned trajectories, the variance estimated from simulation data exhibits a linear time dependence (see Fig.~\ref{Fig_pinning_pdf}(b)), which is characteristic of normal diffusion. Assuming that diffusion and pinning are independent processes, Eq.~(\ref{app_eq_var_traj}) for the DW position variance $\sigma_{\text{DW}}^2(t)$ can be rewritten as:
\begin{equation}
\begin{split}
\sigma_{\text{DW}}^2(t)& = Q(t)Dt + \\
&(1 - Q(t))\left[ \left<\delta X^2\right>_t + Q(t)\left( vt - \left<X\right>_t \right)^2\right],
\end{split}
\label{app_eq_var_traj_dif}
\end{equation}
where $D$ represents the diffusion coefficient of the unpinned motion. Introducing the dimensionless diffusion coefficient $\mathcal{D} = D/(v\Lambda)$, the final expression for $\sigma_{\text{DW}}^2$ is given by:
\begin{equation}
\begin{split}
\frac{\sigma_{\text{DW}}^2(\tau)}{(1-p_0)\Lambda^2} = {}&1 + p_0 - (1 - p_0) e^{-2\tau} - \\
&\left(\left(2 - \mathcal{D}\right)\tau + 2p_0\right) e^{-\tau}.
\end{split}
\label{app_eq_var_final}
\end{equation}
The dependence of DW position variance $\sigma_{\text{DW}}^2(\tau)$ on $\tau$ for various values of parameters $D$ and $p_0$ is shown in Fig.~\ref{Fig_pinning_pdf}(c). The asymptotic expansions of $\sigma_{\text{DW}}^2$ across different time regimes yield

\begin{equation}
\frac{\sigma_{\text{DW}}^2(\tau)}{(1-p_0)\Lambda^2} \approx \begin{cases} 
\mathcal{D}\tau, & \text{if } \tau \ll \mathcal{D}/\tilde{p}_0, \\ 
\tilde{p}_0\tau^2, & \text{if }\mathcal{D}/\tilde{p}_0 \ll \tau \ll 3\tilde{p}_0, \\ 
\frac{1}{3}\tau^3, & \text{if }3\tilde{p}_0 \ll \tau \ll 1, \\ 
1 + p_0, & \text{if } \tau \gg 1.
\end{cases}
\label{app_eq_asymptotics_final}
\end{equation}
where $\tilde{p}_0=p_0+\sqrt{\mathcal{D}}$.

\bibliography{inhomogeneity}
\end{document}